\documentclass[sigconf,screen]{acmart}

\def\fullversion

\setcopyright{acmcopyright}
\copyrightyear{2018}
\acmYear{2018}
\acmDOI{XXXXXXX.XXXXXXX}

\usepackage{booktabs}   
\usepackage{subcaption} 
\usepackage{caption}
\usepackage{colortbl}
\usepackage{bigstrut}
\usepackage{microtype}
\usepackage{tabularx}
\usepackage{multirow}
\usepackage{multicol}
\usepackage{rotating}
\usepackage{lipsum}
\usepackage{balance}
\usepackage{mfirstuc}
\usepackage{tcolorbox}
\usepackage{titlecaps}
\usepackage{listings}
\usepackage{float}
\usepackage[rightcaption]{sidecap}

\usepackage{xcolor}
\usepackage[font=small,labelfont=bf,skip=2pt]{caption}
\usepackage{float}

\def\fullversion
\usepackage{etoolbox}
\newcommand{\ifconference}[1]{{{\ifx\fullversion\undefined{#1}\fi}\xspace}}
\newcommand{\iffullversion}[1]{{{\ifx\conference\undefined{#1}\fi}\xspace}}

\usepackage{graphicx}  
\usepackage{lipsum}  
\newcommand{\hide}[1]{} 
\usepackage{xspace}
\usepackage{textcomp}
\usepackage{comment} 
\usepackage{verbatim}
\usepackage{dblfloatfix}
\usepackage{afterpage}

\definecolor{mypink1}{rgb}{0.858, 0.188, 0.478}
\definecolor{mypink2}{RGB}{219, 48, 122}
\definecolor{mypink3}{cmyk}{0, 0.7808, 0.4429, 0.1412}
\definecolor{mygray}{gray}{0.6}
\newcommand{\red}[1]{{\textcolor{red}{#1}}}

\usepackage{url}

\newcommand{\defn}[1]{\emph{\textbf{\boldmath #1\unboldmath}}} 
\newcommand{\emp}[1]{\emph{\textbf{#1}}} 
\newcommand{\fname}[1]{\textsf{#1}} 
\newcommand{\vname}[1]{{\mathit{#1}}} 
\newcommand{\mathfunc}[1]{\mathit{#1}}

\usepackage[ruled,lined,linesnumbered,noend]{algorithm2e}
\usepackage[noend]{algpseudocode}

\SetAlCapNameFnt{\small}
\SetAlCapFnt{\small}

\makeatletter
\patchcmd{\@algocf@start}
  {-1.5em}
  {0pt}
  {}{}
\newcommand{\nosemic}{\renewcommand{\@endalgocfline}{\relax}}
\newcommand{\dosemic}{\renewcommand{\@endalgocfline}{\algocf@endline}}

\SetSideCommentLeft
\SetKwInput{notations}{Notations}
\SetKwInput{notes}{Notes}
\SetKwInput{maintains}{Maintains}

\SetKwProg{myfunc}{Function}{}{}
\SetKwFor{parForEach}{ParallelForEach}{do}{endfor}
\SetKwFor{Justrepeat}{Repeat}{}{}

\SetKw{MIN}{min}
\SetKw{MAX}{max}
\SetKw{OR}{or}
\SetKw{AND}{and}

\SetCommentSty{mycommfont}

\usepackage{cleveref}
\crefname{section}{Sec.}{Sec.}
\crefname{theorem}{Thm.}{Thm.}
\crefname{lemma}{Lem.}{Lem.}
\crefname{corollary}{Col.}{Col.}
\crefname{table}{Tab.}{Tab.}
\crefname{algorithm}{Alg.}{Alg.}
\crefname{figure}{Fig.}{Fig.}
\crefname{fact}{Fact}{Fact}
\Crefname{table}{Tab.}{Tab.}
\crefname{problem}{Problem}{Problem}

\newtheorem{theorem}{Theorem}[section]
\newtheorem{lemma}[theorem]{Lemma}
\newtheorem{corollary}[theorem]{Corollary}

\let \originalleft \left
\let\originalright\right
\renewcommand{\left}{\mathopen{}\mathclose\bgroup\originalleft}
\renewcommand{\right}{\aftergroup\egroup\originalright}

\usepackage{scalerel} 

\newtheoremstyle{exampstyle}
{.5em} 
{1em} 
{\it} 
{.5em} 
{\it \bfseries} 
{.} 
{.5em} 
{} 
\theoremstyle{exampstyle} 
\theoremstyle{exampstyle} 
\theoremstyle{exampstyle} 
\theoremstyle{exampstyle} 

\makeatletter
\renewenvironment{proof}[1][\proofname]{\par
\vspace{-2\topsep}
\pushQED{\qed}%
\normalfont
\topsep0pt \partopsep0pt 
\trivlist
\item[\hskip\labelsep
      \itshape
  #1\@addpunct{.}]\ignorespaces
}{%
\popQED\endtrivlist\@endpefalse
}

\newcommand{\whp}[1]{\emph{whp}}

\newcommand{\true}{\emph{true}}
\newcommand{\false}{\emph{false}}

\usepackage{pifont}

\newcommand{\modelop}[1]{\texttt{#1}}
\newcommand{\forkins}{\modelop{fork}}

\newcommand{\thread}{thread}

\usepackage[shortlabels]{enumitem}

\setlist{topsep=0.3em,itemsep=0.2em,parsep=0.1em,leftmargin=*}

\usepackage{float}
\usepackage[labelfont=bf,font={small},aboveskip=0em, belowskip=0em]{caption}

\usepackage[labelfont=bf,list=true,skip=0em]{subcaption}
\usepackage[rightcaption]{sidecap}

\usepackage{wrapfig}

\usepackage{array}
\newcolumntype{L}[1]{>{\raggedright\let\newline\\\arraybackslash\hspace{0pt}}m{#1}}
\newcolumntype{C}[1]{>{\centering\let\newline\\\arraybackslash\hspace{0pt}}m{#1}}
\newcolumntype{R}[1]{>{\raggedleft\let\newline\\\arraybackslash\hspace{0pt}}m{#1}}
\newcolumntype{B}{>{\bf}c}
\usepackage{rotating}

\usepackage{booktabs} 
\usepackage{multicol,multirow}
\usepackage{longtable} 
\usepackage{supertabular} 
\usepackage{colortbl}
\usepackage{bigstrut}

\usepackage{titlesec}
\titleformat{\subsection}{\normalfont\large\bfseries}{\thesubsection}{1em}{}

\titlespacing*{\section}{0pt}{0.3em}{0.2em} 
\titlespacing*{\subsection}{0pt}{0.3em}{0.2em} 
\titlespacing*{\subsubsection}{0pt}{0.1em}{1em} 
\newcommand{\mysubsubsection}[1]{{#1}.}
\titleformat{\subsubsection}[runin]
{\normalfont\normalsize\bfseries}{\thesubsubsection}{1em}{\mysubsubsection}

\newcommand{\myparagraph}[1]{\vspace{.1em}\noindent\emp{#1}\enspace}

\usepackage{mdframed}
\definecolor{framelinecolor}{RGB}{68,114,196}
\mdfdefinestyle{mystyle}{linecolor=framelinecolor,innertopmargin=1pt,innerbottommargin=2pt,backgroundcolor=gray!20,skipabove=2pt,skipbelow=0pt}
\mdfdefinestyle{densestyle}{linecolor=framelinecolor,innertopmargin=0,innerbottommargin=0,leftmargin=0,rightmargin=0,backgroundcolor=gray!20}
\mdfdefinestyle{compactcode}{linecolor=framelinecolor,innertopmargin=1pt,innerbottommargin=1pt,backgroundcolor=gray!20,skipabove=0pt,skipbelow=0pt,leftmargin=0,rightmargin=0}

\usepackage{framed}

\usepackage{listings}

\newdimen\zzsize
\newdimen\kwsize
\newcommand{\basicstyle}{\fontsize{\zzsize}{1\zzsize}\ttfamily}
\newcommand{\keywordstyle}{\fontsize{\kwsize}{1\kwsize}\ttfamily\bf}

\newdimen\zzlstwidth
\usepackage{tikz} 

\setcopyright{none}
\renewcommand\footnotetextcopyrightpermission[1]{} 

\newcommand{\degree}{d}
\newcommand{\dmin}{k}
\newcommand{\dbar}{d_{\text{\it avg}}}
\newcommand{\nei}{N}
\newcommand{\degreestar}{\tilde{d}}
\newcommand{\induceddegree}{induced degree}
\newcommand{\kmax}{k_{\max}}

\newcommand{\kcore}{$k$-core\xspace}
\newcommand{\atominc}{\fname{atomic\_inc}}
\newcommand{\atomdec}{\fname{atomic\_dec}}
\newcommand{\subround}{subround\xspace}
\newcommand{\HBS}{HBS\xspace}

\newcommand{\bagput}{\mf{BagInsert}\xspace}

\newcommand{\bagpack}{\mf{BagExtractAll}\xspace}

\newcommand{\coreness}{\mbox{\red{coreness}}\xspace}

\newcommand{\mf}[1]{{\mbox{\sc{#1}}}}

\newcommand{\truecoreness}{{\kappa}\xspace}
\newcommand{\maxcoreness}{{k_{\max}}\xspace}

\newcommand{\frontier}{{\mathcal{F}}\xspace}
\newcommand{\nextfrontier}{\mathcal{F}_{\text{next}}}
\newcommand{\alive}{active}
\newcommand{\alivevertices}{{\mathcal{A}}\xspace}

\newcommand{\FPeel}{\mf{Peel}\xspace}

\newcommand{\FError}{\mf{Validate}\xspace}
\newcommand{\FInitSampler}{\mf{SetSampler}\xspace}
\newcommand{\FSetSampler}{\mf{Resample}\xspace}

\newcommand{\FHistogram}{\mf{Histogram}\xspace}
\newcommand{\pack}{\mf{Pack}}
\newcommand{\FBuildBuckets}[1]{\mf{BuildBuckets(}#1\mf{)}\xspace}
\newcommand{\FGetNextBucket}{\mf{GetNextBucket()}\xspace}
\newcommand{\FUpdate}[1]{\mf{DecreaseKey(}#1\mf{)}\xspace}

\newcommand{\FGetBucket}{\mf{GetBucket}\xspace}

\newcommand{\customIf}[2]{%
    \textbf{if} #1 \textbf{then} #2 
}
\newcommand{\customForEach}[2]{%
    \textbf{foreach} #1 \textbf{do} #2
}

\newcommand{\exphits}{\mathfunc{\mu}\xspace}

\newcommand{\reducerate}{r}
\newcommand{\sampler}{\sigma}
\newcommand{\countingbag}{\mathcal{C}}

\newcommand{\mode}{\text{\it mode}}
\newcommand{\rate}{\text{\it rate}}
\newcommand{\cnt}{\text{\it cnt}}
\newcommand{\cntproof}{s}
\newcommand{\rateproof}{p}

\newcommand{\PKC}{\textsf{PKC}\xspace}
\newcommand{\pkc}{\PKC}
\newcommand{\ParK}{\textsf{ParK}\xspace}
\newcommand{\Park}{\ParK}
\newcommand{\park}{\ParK}
\newcommand{\Julienne}{\textsf{Julienne}\xspace}
\newcommand{\julienne}{\Julienne{}}
\newcommand{\GBBS}{\textsf{Julienne}\xspace}
\newcommand{\BZ}{\textsf{BZ}\xspace}

\newcommand{\OK}{\emph{OK}\xspace}

\newcommand{\TW}{\emph{TW}\xspace}

\newcommand{\EH}{\emph{EH}\xspace}
\newcommand{\SD}{\emph{SD}\xspace}
\newcommand{\CW}{\emph{CW}\xspace}
\newcommand{\HL}{\emph{HL14}\xspace}
\newcommand{\HLs}{\emph{HL12}\xspace}

\newcommand{\NA}{\emph{NA}\xspace}

\newcommand{\EU}{\emph{EU}\xspace}

\newcommand{\GL}{\emph{GL5}\xspace}

\newcommand{\COS}{\emph{COS5}\xspace}
\newcommand{\TRCE}{\emph{TRCE}\xspace}
\newcommand{\BBL}{\emph{BBL}\xspace}
\newcommand{\GRD}{\emph{GRID}\xspace}
\newcommand{\CBC}{\emph{CUBE}\xspace}
\newcommand{\HCNS}{\emph{HCNS}\xspace}
\newcommand{\HPL}{\emph{HPL}\xspace}
\newcommand{\knn}{$k$-NN}

\newcommand{\GLsfull}{\emph{GeoLife10}\xspace}

\newcommand{\KNN}{$k$-NN\xspace}

\newcommand{\yihan}[1]{{\color{blue}{\bf Yihan:} #1}}

\newcommand{\youzhe}[1]{{\color{cyan}{\bf Youzhe:} #1}}

\newcommand{\revise}[1]{{\color{black} #1}}

\begin{document}

\makeatletter
\gdef\@copyrightpermission{
  \begin{minipage}{0.2\columnwidth}
   \href{https://creativecommons.org/licenses/by/4.0/}{\includegraphics[width=0.90\textwidth]{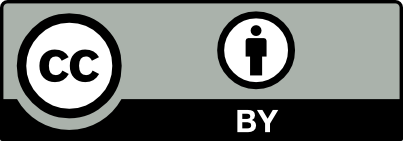}}
  \end{minipage}\hfill
  \begin{minipage}{0.8\columnwidth}
   \href{https://creativecommons.org/licenses/by/4.0/}{This work is licensed under a Creative Commons Attribution International 4.0 License.}
  \end{minipage}
  \vspace{5pt}
}
\makeatother

\title{Parallel $k$-Core Decomposition: Theory and Practice}

\settopmatter{authorsperrow=4}
\author{Youzhe Liu}
\email{yliu908@ucr.edu}
\affiliation{%
  \institution{UC Riverside}
  \city{}
  \country{}
}
\author{Xiaojun Dong}
\email{xdong038@ucr.edu}
\affiliation{%
  \institution{UC Riverside}
  \city{}
  \country{}
}
\author{Yan Gu}
\email{ygu@cs.ucr.edu}
\affiliation{%
  \institution{UC Riverside}
  \city{}
  \country{}
}
\author{Yihan Sun}
\email{yihans@cs.ucr.edu}
\affiliation{%
  \institution{UC Riverside}
  \city{}
  \country{}
}

\renewcommand{\shortauthors}{Liu et al.}

\begin{abstract} This paper proposes efficient solutions for \kcore{} decomposition with high parallelism. 
    The problem of \kcore{} decomposition is fundamental in graph analysis and has applications across various domains. 
    However, existing algorithms face significant challenges in achieving work-efficiency in theory and/or high parallelism in practice, 
    and suffer from various performance bottlenecks.

    We present a simple, work-efficient parallel framework for $k$-core decomposition that is easy to implement and adaptable to various strategies for improving work-efficiency. 
    We introduce two techniques to enhance parallelism: a sampling scheme to reduce contention on high-degree vertices, and vertical granularity control (VGC) to mitigate scheduling overhead for low-degree vertices. 
    Furthermore, we design a hierarchical bucket structure to optimize performance for graphs with high coreness values.
    
    We evaluate our algorithm on a diverse set of real-world and synthetic graphs. 
    Compared to state-of-the-art parallel algorithms, including \ParK, \PKC, and \Julienne, 
    our approach demonstrates superior performance on 23 out of 25 graphs when tested on a 96-core machine. 
    Our algorithm shows speedups of up to 315$\times$ over \Park{}, 33.4$\times$ over \PKC{}, and 52.5$\times$ over \Julienne{}.
    \let\thefootnote\relax\footnotetext{This paper was accepted at the International Conference on Management of Data (SIGMOD 2025).}
	\setcounter{footnote}{0}
    \end{abstract}



\fancyhead{} 

\maketitle

\section{Introduction}\label{sec:intro}
Analyzing real-world graphs is crucial in a wide range of applications.
One of the most widely-used techniques for identifying densely connected regions in a network is the $k$-core decomposition~\cite{seidman1983network,matula1983smallest}.
This problem about subgraph structure detection finds applications in various fields, 
including social network analysis~\cite{kong2019k, zhang2017finding, kitsak2010identification,malliaros2020core, boguna2004models, giatsidis2011evaluating}, 
risk assessment~\cite{kong2019k, morone2019k, garcia2017ranking, malliaros2020core, burleson2020k, zhang2010using}, 
bioinformatics~\cite{kong2019k, wuchty2005peeling, malliaros2020core, cheng2013local, emerson2015k}, 
and system robustness analysis~\cite{kong2019k, malliaros2020core, burleson2020k, yao2022study, sun2020new}.

\begin{figure}[t]
  \centering
  \includegraphics[width=0.7\columnwidth]{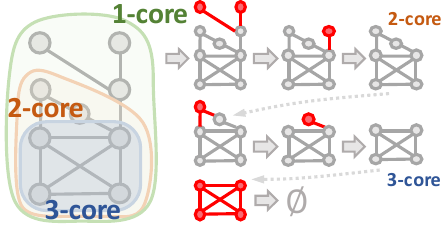}
  \caption{\textbf{An example of $k$-core decomposition with \boldmath $\maxcoreness = 3$.\unboldmath}
  Vertices and edges peeled in each subround are marked as red.}\label{fig:kcore_eg}
\end{figure}

Given a graph $G = (V, E)$ with $n=|V|$ vertices and $m=|E|$ edges, the $k$-core of $G$ is the maximal subgraph $G' = (V', E')$ of $G$ in which every vertex has degree at least $k$ (see an illustration in \cref{fig:kcore_eg}).
The $k$-core decomposition of a graph $G$ identifies a sequence of non-empty subgraphs $G_0, G_1, \ldots, G_{k_{\max}}$ for all possible $k$ values, where $G_i$ is the $i$-core of $G$. 
The \defn{coreness} of a vertex, denoted as $\truecoreness[v]$, is the maximum value of $k$ such that $v$ is in $G_k$. 
The coreness of a graph, denoted as $\kmax$, is the maximum coreness among all vertices. 
The output of the $k$-core decomposition is the coreness for each vertex, which can be used to reconstruct any $G_i$. 

\begin{figure}[t!]
  \centering
  \includegraphics[width=\columnwidth]{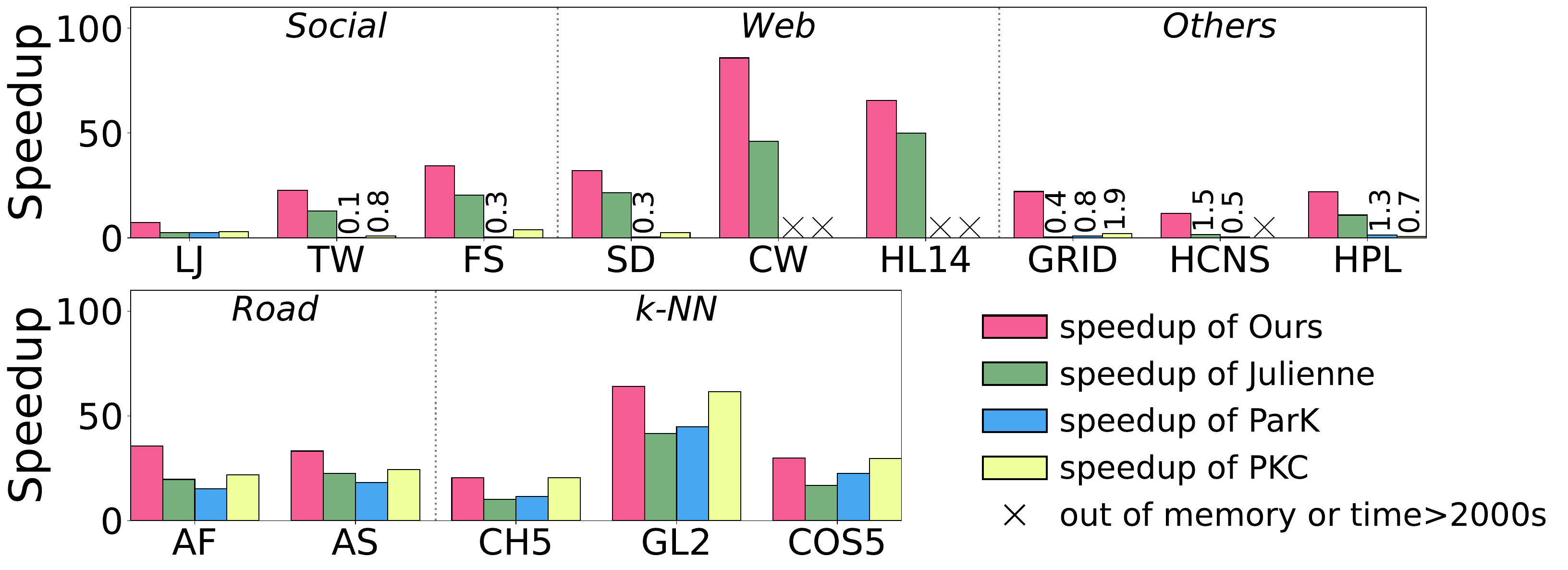}
  \caption{\small \textbf{Speedup of \ParK~\cite{dasari2014park}, \PKC~\cite{kabir2017parallel}, \Julienne~\cite{dhulipala2017, gbbs2021}, and our algorithm, 
  over to the best sequential time (our sequential time or the BZ algorithm time~\cite{batagelj2003m}) on 14 representative graphs. Higher is better.} 
  Full results are in \cref{table:fulltable}.
  Numbers below 2 are given on the bars, meaning the parallel code is no more than $2\times$ faster than a sequential one.
  }\label{fig:overall_comp_seq}
\end{figure}

Consider the ever-growing size of today's real-world graphs, 
which can reach terabyte scale,
it is essential to consider parallelism in the \kcore{} computation. 
In this paper, we focus on the high-performance algorithm to compute exact $k$-core decomposition with high parallelism.
While \kcore{} is simple in the sequential setting, and various efficient solutions exist~\cite{seidman1983network,batagelj2003m}, 
parallel $k$-core is a notoriously hard problem. 
Starting from $k=0$, the sequential algorithm keeps removing vertices with degree no more than $k$ and update their neighbors' degrees, until all nodes have a degree at least $k+1$, obtaining the $(k+1)$-core. Then the algorithm will increment $k$ and repeat this process. 
An example of this process, usually referred to as the ``peeling'' process, is given in \cref{fig:kcore_eg}.
This sequential approach requires $O(n+m)$ number of operations (aka.\ work or time complexity). 
However, parallelizing this sequential method is highly non-trivial.
Despite \kcore{} is studied in many papers and implemented in many libraries~\cite{gonzalez2014graphx,gonzalez2012powergraph,shun2013ligra,dhulipala2017,kabir2017parallel, cheng2011efficient,montresor2011distributed, li2021k,dasari2014park, esfandiari2018parallel, liu2024parallel, mehrafsa2020vectorising,tripathy2018scalable,zhao2024speedcore, ahmad2023accelerating, konduri2022implementation, khaouid2015k,zhang2017accelerating, zhao2024pico, wen2018efficient}, 
many challenges remain, both in theory and in practice, to achieve a parallel \kcore{} algorithm that is simple, efficient, and scalable on various types of graphs.  
For instance, in \cref{fig:overall_comp_seq}, we show that on a 96-core machine, each state-of-the-art parallel \kcore solution can be slower than a sequential implementation on certain graphs, and the ``worst cases'' vary significantly between algorithms. 
This highlights the intricacies of optimizing the \kcore{} algorithm in the parallel setting. 

\textbf{In this paper, we propose a parallel \kcore{} algorithm with a series of novel algorithmic techniques. 
Our algorithm is efficient, practical, and achieves high performance across various graph types.}
In the following, we briefly overview the challenges in existing work, and our solutions to overcome them. 

\myparagraph{Theoretical challenges and our contributions.} 
The probably most surprising challenge we have observed is that, we are unaware of any analysis on work-efficiency\footnote{The \emph{work} of a parallel algorithm is its time complexity running on one core. A parallel algorithm is \emph{work-efficient} if its work is the same as the best sequential time complexity.} for the existing parallel implementations.
Most of existing parallel \kcore{} decomposition algorithms have $O(m+\maxcoreness n)$ work, 
rendering a significant overhead caused by parallelism. 
The only known work-efficient parallel \kcore{} algorithm is from \Julienne~\cite{dhulipala2017}.
Unfortunately, this algorithm is mainly of theoretical interest. 
During the entire peeling process, it maintains a \emph{bucketing structure}, 
which maps each possible degree $d$ to all vertices with degree $d$.  
However, maintaining the full mapping incurs
performance overheads and coding complexity in practice. 
Hence, their implementation~\cite{dhulipala2017} did not adopt the bucketing structure as described 
and used a simplified alternative, which only maintains at most $b$ buckets at any time. 
It remains unknown if their simplified implementation is theoretically efficient. 

In this paper, we present a surprisingly simple parallel \kcore algorithm, shown in \cref{algo:framework}, with $O(n+m)$ work. 
Our framework does not need the complicated bucketing structure to achieve work-efficiency, making it simple to implement. 
Our analysis indicates that the simplified implementation of \Julienne{} is also work-efficient, 
and some other existing algorithms with $O(m+\maxcoreness n)$ work can be made work-efficient with minor modifications. 


\myparagraph{Practical challenges and our contributions.} While our analysis indicates that achieving work-efficiency in \kcore{} algorithms is not hard,
translating this into a fast, practical implementation is challenging due to the difficulty of achieving high parallelism.
Existing implementations perform the peeling process in either an online or an offline manner. 
When peeling vertices with degree~$k$, the offline approach, implemented in \Julienne{}~\cite{dhulipala2017}, 
chooses to collect all vertices that require degree decrement in a batch, counts the number of occurrences for each of them,
and applies all of them in parallel. 
The benefit of the algorithm is race-freedom. 
However, since this approach is fully synchronous, meaning that the degree decrement for each batch (which we call a \defn{subround} in this paper) must fully finish before the next batch starts, causing high overhead to synchronize threads between subrounds. 
One adversarial example is a $\sqrt{n}\times \sqrt{n}$ grid (the \GRD{} graph in \cref{fig:overall_comp_seq}), 
that incurs $O(\sqrt{n})$ subrounds.
An illustration of this process is given in \cref{fig:localsearch}(a).
On this graph, \Julienne{} is slower than a sequential implementation due to high scheduling overhead. 

Many other parallel \kcore{} implementations, such as \pkc{}~\cite{kabir2017parallel} and \park{}~\cite{dasari2014park}, 
use the \emph{online} approach. 
When peeling a vertex $v$, the degrees of its neighbors are directly decremented via atomic operations.
The online algorithm follows a simple framework, and 
certain optimizations can also be used to reduce the synchronization cost~\cite{kabir2017parallel}. 
However, for high-degree vertices, a large number of concurrent decrements can cause heavy contention, 
which degrades performance.
As a result, \pkc{} and \park{} are slower than sequential algorithms on dense networks
(e.g., \TW{} and \SD{} in \cref{fig:overall_comp_seq}).  

To resolve these challenges, we propose several novel techniques. 
We adopt the online framework with two additional techniques, given in \cref{sec:online-framework}. 
The first is a \emph{sampling} scheme to reduce concurrent decrements on high-degree vertices.
For a vertex $v$ with a large degree, 
we select a subset of its neighbors, 
and only the removal of these neighbors decrements $v$'s degree.
The key challenge of sampling is to accurately estimate the true degree of each vertex $v$, 
and to identify $v$ when it is ready to be peeled. 
Our analysis ensures the sampling algorithm provides correct estimates with high probability.
On dense graphs, sampling improves the performance of our code by up to 4.31$\times$. 

Our second technique is to reduce the high synchronization cost between peeling subrounds. 
By using the online framework, our algorithm allows asynchronous degree decrements.
When we peel a low-degree vertex $v$, the actual computation to process all its neighbors is minimal,
and the scheduling overhead to create and synchronize this thread may dominate the cost.
To address this,
we propose a \emph{local search} algorithm to reduce the number of subrounds and hide scheduling overhead. 
This technique improves the performance of our code by up to 31.2$\times$. 

Finally, we propose a new hierarchical bucketing structure (HBS) to further improve performance in \cref{sec:bucketing}.
Instead of using no bucket structure (or equivalently, using a single bucket) as in our proposed framework, 
or a fixed number of buckets as in \Julienne{}, 
HBS efficiently manages vertices in each round. 
On average (geometric mean across all graphs), our new design improves the performance by 1.6$\times$ compared to the plain version (no-bucket structure),
and is $1.3\times$ faster than using 16 buckets. 
\hide{
On the other hand, maintaining bucketing structures can incur high overhead for graphs with a small $\maxcoreness$. 
In our implementation, we start with no bucketing structure, and switch to maintaining a number of buckets once the current peeling round $k$ reaches a threshold (8 in our code). Instead of organizing vertices based on their exact degrees, we will maintain $O(\log n)$ buckets, the $i$-th bucekts maintaining vertices with degree in range $[k+2^{i-1},k+2^{i})$. 
In this way, we avoid frequently initializing the buckets and moving elements across buckets, which leads to significant performance improvement on graphs with large corenesses. \yihan{maybe mention some experimental results}
}

With our new framework and techniques, our algorithm achieves high performance in experiments. 
We compare our implementation against three state-of-the-art parallel implementations: \Julienne{}, \Park{} and \PKC{}, 
testing a diverse set of graphs, including social networks, web graphs, road networks, \KNN graphs, and other graphs including synthetic graphs and simulation graphs. 
Due to different designs, each baseline has lower performance than a sequential implementation on certain graphs. 
Our algorithm consistently performs well on all types of graphs, 
outperforming the best sequential algorithm by 7.3--84$\times$.
On dense graphs, 
our algorithm is faster than all baselines, 
with 1.14--3.17$\times$ faster than \Julienne{}, 2.6--315$\times$ faster than \Park{}, and 1.5--28$\times$ faster than \PKC{}.
On most sparse graphs, our algorithm outperforms the baselines, except for two cases where it is within 12\% of the best baseline.
On sparse graphs, our algorithm is up to 53$\times$ faster than \Julienne{}, 28$\times$ faster than \Park{}, and 11$\times$ faster than \PKC{}. 
We also carefully evaluated the performance gain from the three new techniques in experiments. 

We release our code in~\cite{kcorecoderelease}. \ifconference{For page limit, we provide the full version of our paper in the supplementary material. }

\section{Preliminaries}\label{sec:prelim}

\myparagraph{Notations.}
Given an undirected graph $G=(V,E)$, 
the $k$-core of a graph $G$, noted as $G_k$, is the largest subgraph of $G$ where every vertex has degree at least $k$.
The \defn{coreness} of a vertex, noted as $\truecoreness(v)$, is the maximum value of $k$ such that $v$ is in $G_k$. 
The coreness of a graph, noted as $\kmax$, is the maximum coreness among all vertices. 
The \defn{\kcore{} decomposition} of $G$ identifies the sequence of subgraphs $G_0, G_1, \dots, G_{\kmax}$. 
With clear context, we use \kcore{} to refer to \kcore{} decomposition. 
Our algorithm computes the coreness for each vertex $v \in V$, 
which can be used to recover any $G_i$. 

Most \kcore{} algorithms are based on the \emph{peeling} process, 
where vertices with the smallest degrees are removed, and their neighbors' degrees are updated accordingly.
We use $\degree(v)$ to represent the degree of a vertex in the original graph. 
To distinguish from the original degree, 
we use \defn{\induceddegree{}} to denote the degree of~$v$ during the peeling process, 
which may be decremented by the removal of its neighbors, and denote it as $\degreestar[v]$ for vertex $v$. 

We say $O(f(n))$ {with high probability (\whp{}) in $n$} to indicate $O(cf(n))$ with probability at
least $1-n^{-c}$ for any $c \geq 1$.  
We say an event happens {with high probability (\whp{}) in $n$} to indicate that probability is at
least $1-n^{-c}$ for any $c \geq 1$ as a parameter.  
When clear from context we drop the ``in $n$''.

\myparagraph{Parallel Computational Model.}
\revise{We use the work-span model in the classic multithreaded model with binary fork-join~\cite{blelloch2020optimal}.
We assume a set of \thread{}s that share a common memory. }
A process can \forkins{} two child software \thread{s} to work in parallel.
When both children complete, the parent process continues.
The \defn{work} of an algorithm is the total number of instructions.
An algorithm is \defn{work-efficient} if its work is asymptotically the same as the best sequential algorithm. 
\revise{
  The \defn{span} of an algorithm is the length of the longest dependence chain among operations.
We can execute a computation with $W$ work and $S$ span with using a randomized work-stealing scheduler~\cite{BL98,arora2001thread} in time $W/P+O(S)$ \whp{} in $W$. 
}

\revise{
  A major goal in our work is to reduce the synchronization cost in the \kcore{} algorithm. 
  Unfortunately, this cost is not well-reflected by the classic work-span model. 
  More precisely, while each fork/join operation is counted as a constant cost in theory, such a constant is usually large due to the scheduling scheme. 
  To formally analyze the cost, we borrow the concept of \defn{burdened span} from Cilkview~\cite{he2010cilkview}, a profiling tool
  for multicore programs. 
  Burdened span charge a cost of $\omega$ for a fork/join operation to reflect scheduling overhead, and a unit cost as usual for other operations. 
  We use the default parameter $\omega=15,000$ from the original Cilkview paper. 
  We will show that our proposed technique improves the burdened span compared to existing solutions. 
  We also use Cilkview~\cite{he2010cilkview} to measure the exact burdened span on real-world graphs in our experiments.}

We use atomic operations \atominc{}$(p)$ and \atomdec{}$(p)$, which atomically reads the memory location pointed to by $p$, and increment (or decrement) the integer value stored at $p$ by 1. The function returns the original value stored at $p$. 
We assume constant work for each atomic operation in theoretical analysis.
We note that, however, the cost for atomic operations can usually be large in practice, especially when many threads are concurrently
updating the same memory location. 
One effort in our paper is to avoid such high contention by reducing the number of atomic updates to the same memory location. 
\revise{  
  To capture this, we define \defn{contention}~\cite{acar2017contention} experienced by an operation as the maximum number of operations that may concurrently modify the same memory location.   The contention of an algorithm is the highest contention among all operations in the algorithm. In \cref{sec:sampling}, we will show that our sampling technique effectively reduces the contention. 
}




\begin{table}[t]
  \centering
  \small
  \begin{tabular}{|cl|}
    \multicolumn{2}{@{}l}{Notations for the input graph:}\\
    \hline
    $G = (V, E)$ & Graph $G$ with vertex set $V$ and edge set $E$ \\
    $n, m$ & $n=|V|,m=|E|$ \\
    $\truecoreness(v)$ & The coreness of vertex $v$ \\
    $\maxcoreness$ & The maximum coreness value in the graph \\
    $\nei(v)$ & The set of neighbors of vertex $v$ \\
    $\degree(v)$ & The degree of vertex $v$ \\
    \hline
    \multicolumn{2}{@{}l}{Notations used in the algorithm:}\\
    \hline
    $\mathcal{A}$ & The set of active vertices (not peeled yet) \\
    $\mathcal{F}$ & The set of frontier vertices \\
    $\degreestar[v]$ & The induced degree of $v$ during the algorithm\\ 
    $k$ & The current peeling round\\
    \multicolumn{2}{|c|}{Notations used in sampling are defined in \cref{algo:sample_func}}\\
    \hline
  \end{tabular}
  \caption{\bf Notation used in this paper.}
  \label{tab:notation}

\end{table}

\myparagraph{Parallel Primitives.}
The \pack{} function is a widely-used parallel primitive in this paper. 
Given an array $A$ of elements and a predicate function $f$, 
the \emph{pack} function returns all elements in $A$ that satisfy $f$ in a new array. 
This function takes $O(|A|)$ work and can be highly parallelized. 
\Julienne{} uses the \FHistogram{} function, which takes an array $A$ of keys, and count the occurrences of each key in $A$.
It can be computed efficiently using parallel semisort~\cite{gu2015top,dong2023high}.

\myparagraph{Parallel Hash Bag.} 
In this paper, we use a data structure \emph{parallel hash bag}~\cite{dong2021efficient,wang2023parallel}. 
A parallel hash bag supports concurrent insertion and resizing, and extracting all elements into a consecutive array. 
More precisely, a parallel hash bag supports two operations:
\begin{itemize}[leftmargin=*,topsep=0pt, partopsep=0pt,itemsep=0pt,parsep=0pt]
  \item \bagput{}($v$): (concurrently) add the element $v$ into the bag.  
  \item \bagpack{}(): extract all elements in the bag into an array. 
\end{itemize}

Hashing is a widely-used technique for accelerating parallel data access~\cite{shun2014phase, Alcantara2009, dong2024parallel, ding2023efficient}.
Similar to a hash table~\cite{shun2014phase}, a hash bag maintains a set of elements by hashing
every element into certain index, and resolve conflicts by linear probing. 
A hash bag is initialized as an array with $O(n)$ slots, 
where $n$ is the maximum possible number of elements appearing in the bag at the same time. 
The array is conceptually divided into chunks of sizes $\lambda, 2\lambda, 4\lambda, ...$ ($\lambda=2^8$ in implementation).
At the beginning, elements are inserted into the chunk of size $\lambda$.
Once the current chunk reaches a desired load factor, 
we move onto the next chunk of size $2\lambda$, and so on so forth. 
The \bagpack{}() function only needs to consider the prefix chunks that have been used instead of the entire array.
Therefore, the complexity of \bagpack{}() is $O(\lambda+t)$, where $t$ is the number of elements in the hash bag.
In this paper, we use parallel hash bag to maintain the frontiers and some vertex sets in our algorithm. 

\myparagraph{Bucketing Structure for \kcore{} Algorithms.} 
The bucketing structure organizes data with integer keys and supports efficient updates and finding the element with the smallest key. 
A parallel bucketing structure has been proposed by~\cite{dhulipala2017} to support the \kcore{} algorithm in \Julienne{}.
The design of the bucketing structure has been widely studied \cite{li2013parallel, shi2021parallel, shi2023theoretically,dhulipala2017,gbbs2021} and has many applications. 
In \cref{sec:bucketing}, we present more details about the existing bucketing structure and our new design, which not only enhances the performance of the \kcore algorithm but is also of independent interest.

\section{Algorithm Framework and Design}\label{sec:alg}
In this section, we present our algorithmic framework for $k$-core with theoretical analysis. 
Many existing implementations fit into this framework. 
However, as mentioned, we are unaware of any analysis to prove their work-efficiency. 
For example, both \ParK~\cite{dasari2014park} and \PKC~\cite{kabir2017parallel} proved $O(\maxcoreness n+m)$ time complexity for their algorithms, 
where $\maxcoreness$ can be $O(\sqrt{m})$ in the worst case. 
\Julienne{}~\cite{dhulipala2017} introduced a \kcore{} algorithm with $O(m+n)$ work. 
However, this algorithm is mostly of theoretical interest.
Due to complicated algorithmic details that may cause a performance overhead in practice, 
their open-source code implemented a simpler (and more practical) version instead of the algorithm described in their paper.
It is therefore unclear what the cost is for their implementation. 
This motivates us to consider \emph{whether there exists a simple and work-efficient parallel \kcore{} solution}. 
Interestingly, we observed that a simple algorithmic framework can essentially give $O(m+n)$ work. 
Our results show that the implementation in \Julienne{} is indeed theoretically efficient, as well as a simple variant of \Park{} or \PKC{}. 
\revise{In fact, while many existing algorithms roughly follows the high-level idea of the framework, 
we believe that the formalization of the framework helps enable simple and general analysis for work-efficiency,
reveal the advantages/disadvantages of existing approaches, and inspires the main techniques in this paper. }
In the following, we show the framework in \cref{algo:framework}, describe and analyze it in~\cref{sec:framework}, and finally discuss different strategies for the peeling process in~\cref{sec:peeling}.


\subsection{Framework}\label{sec:framework}

We present our algorithm framework in \cref{algo:framework}.
Given a graph $G=(V,E)$, the algorithm returns the coreness $\truecoreness[v]$ of each $v\in V$.

As with the existing work, our algorithm is also \emph{frontier-based}, 
where vertices with the same degree are peeled in parallel, organized as a \emph{frontier}, denoted as $\frontier$. 
In addition, our algorithm maintains a set of \defn{\alive{} vertices} $\alivevertices$, 
which are the vertices that have not been peeled. 
We call $\alivevertices$ the \defn{\alive{} set}. 
In the peeling process, we use $\degreestar[v]$ to track the \induceddegree{} of $v$. 
We first initialize $\degreestar[v]$ as the degree of $v$ in the input graph, and $\alivevertices$ as the entire vertex set $V$. 
We use $k$ to denote the current peeling round, starting with $k=0$. 
In round $k$, we first initialize the frontier $\frontier$ by extracting the vertices from $\alivevertices$ with \induceddegree{} $k$ (\cref{line:extract}). 
The coreness of all vertices in the frontier can be set to $k$. 
We then peel all vertices in $\frontier$ (\cref{line:peel}). 
For each vertex $v\in \frontier$, peeling it will decrement the \induceddegree{} for each of its neighbors by 1. 
When the \induceddegree{} of any vertex $u$ hits $k$, $u$ is added to a set $\nextfrontier$, which becomes the next frontier. 
We repeat this process until the frontier becomes empty. 
We call each iteration on \cref{line:subround} a \defn{subround}, which deals with one frontier, and call each iteration on \cref{line:round} a \defn{round}, which can contain multiple subrounds with the same coreness value $k$. 
After each round, we update the \alive{} set $\alivevertices$ (\cref{line:pack}) by only keeping vertices with \induceddegree{s} larger than $k$.

\begin{algorithm}[t]
  \small
  \caption{Work-efficient parallel $k$-core framework}
  \KwIn{Graph $G = (V, E)$}
  \KwOut{The coreness for each vertex}
  \label{algo:framework}
  \SetKwFor{parFor}{parallel\_for}{do}{endfor}
  \SetKwInOut{Maintains}{Maintains}
  \medskip
    $\degreestar[\cdot]\gets \degree(\cdot)$ \tcp*[f]{Initialize the induced degree set of all vertices} \\
    $\alivevertices\gets V$\label{line:init-alive}\tcp*[f]{Active set: vertices that have not been peeled} \\
    $k \gets 0$\\
    
    \While{$\alivevertices\ne \emptyset$} { \label{line:round}
      $\frontier \gets \{v~|~v\in\alivevertices,\degreestar[v]= k\}$  \label{line:generate_frontier}\label{line:extract}\tcp*[f]{The initial frontier in round $k$}\\
      \While{$\frontier\ne \emptyset$}{ \label{line:subround}
      \customForEach{$v\in \frontier$}{$\truecoreness[v]\gets k$}\tcp*[f]{Sets coreness to $k$}\\
        $\frontier \gets \FPeel (\frontier,k)$ \label{line:process_bucket}
      }
  
      $\alivevertices \gets \{v~|~v\in\alivevertices,\degreestar[v]>k\}$\label{line:filter_out_alive}\label{line:pack}\tcp*[f]{Refines the \alive{} set}\\
      $k \gets k + 1$   \label{line:increase_k}
    }
    \Return $\truecoreness[\cdot]$ 
  
  \medskip
  \tcc{A sequential version. Parallel versions are discussed in \cref{sec:peeling}.}
  \myfunc{$\FPeel (\frontier,k)$\label{line:peel}}{ 
    Initialize $\nextfrontier\gets \emptyset$ \tcp*[f]{Buffers the next frontier}\\
    \ForEach{$v\in \frontier$} {
        \ForEach{$u\in N(v)$} {
              $\degreestar[u]\gets \degreestar[u]-1$ \label{line:decrease_degree}\\
              \customIf{$\degreestar[u]=k$} 
              {
                Add $u$ to $\nextfrontier$ \
              } \label{line:update_frontier}
        }
    }
    \Return $\nextfrontier$\tcp*[f]{Returns the next frontier}
  }
  \end{algorithm}

  
      
      

  
            
  
          

\hide{The $\FPeel(\frontier,k)$ function can be implemented differently as long as the work is $O(|\frontier|+\sum_{v\in \frontier}{N(v)})$ \youzhe{in parallel} (i.e., the number of incident vertices and edges).
In line 12-19 (add the labels and refs) we show how it can be done sequentially~\cite{seidman1983network,batagelj2003m} with the designated work bound.
We will later show how to parallelize this step either trivially (in \cref{sec:peeling}) or more efficiently (in \cref{sec:online-framework}).}

The cost of the algorithm relies on three parts: 1) \cref{line:extract} that extracts the initial frontier of a round, 2) the \FPeel{} function that processes all neighbors of the frontier, and 3) \cref{line:pack} that refines the \alive{} set $\alivevertices$. 
Next, we will show that, with proper implementations of the three parts, the framework in \cref{algo:framework} has $O(n+m)$ work. 

\begin{theorem}
    \label{thm:work}
        Assuming $O(|\frontier|+\sum_{v\in \frontier}{\degree(v)})$ work for the $\FPeel(\frontier,\cdot)$ function on~\cref{line:process_bucket}, and
        $O(|\alivevertices|)$ work for the functions on \cref{line:extract,line:pack}, where $\alivevertices$ is the input \alive{} set), 
        the total work of \cref{algo:framework} is $O(n+m)$.
\end{theorem}
    \begin{proof}
        The main work of the algorithm comes from two parts: the $\FPeel()$ function, and maintaining the sets $\frontier$ and $\alivevertices$ on 
        \cref{line:extract,line:pack}.
        
        For the $\FPeel{}$ function,
        note that each vertex will appear in exactly one frontier. 
        As we assumed in the theorem, the work of \FPeel{} is proportional to the total number of neighbors of vertices in the frontier.
        Therefore the total work is $\sum_{v\in V} (1+\degree(v))=O(n+m)$.
        
        We now analyze the two functions that generates $\frontier$ on \cref{line:extract} and $\alivevertices$ on \cref{line:pack}. 
        The theorem assumes that the cost for both of them is proportional to the input size $|\alivevertices|$, and thus they have the same asymptotic cost. Let $\alivevertices_i$ be the \alive{} set of round $i$, which contains all vertices with coreness at least $i$. 
        The total cost is:
        \vspace{-.5em}
        \begin{align*}
        \sum_{i=0}^{\maxcoreness} |\alivevertices_i| &= \sum_{i=0}^{\maxcoreness} |\{ v~|~v\in V, \truecoreness[v]\ge i\}|\\
        &=\sum_{v\in V} (1+\truecoreness[v]) \le \sum_{v\in V} 1+\sum_{v\in V}\degree(v) = O(n+m)
        \end{align*}        
        \vspace{-.1em}
        The third step uses the fact that $\truecoreness(v)\le \degree(v)$ for all $v\in V$. 
        \hide{The main work of the algorithm comes from two parts: the $\FPeel()$ function and generating the sets $\alivevertices$ in \cref{line:filter_out_alive} and $\frontier$ in \cref{line:generate_frontier}.
        
        For the $\FPeel()$ function, since every vertex will be peeled exactly once.
        Hence, the total work for this part is $O(n+m)$.
        Now let's mainly consider the work for \cref{line:filter_out_alive}---\cref{line:generate_frontier} has the same work as generating $\alivevertices$ the last time, so it will only double the work and not asymptotically affect the bound.
        Note that when a vertex is alive in the $k$-th round, it must have degree more than $k$.
        Hence, a vertex with degree $d$ can at most appear in $\alivevertices$ in the first $d$ rounds.
        The total size of all active sets $\alivevertices$ in all rounds is also bounded by $O(n+m)$.
        Generating $\alivevertices$ and $\frontier$ only requires to linearly scan over the active sets, leading to the total work as stated.
        }
    \end{proof}
        
The key in the analysis is to bound to total size of all active sets.
Although it does not seem too complicated, to the best of our knowledge, we are unaware of existing work with this result. 

\hide{
Note that this paper is about parallel \kcore.
We will discuss how to parallelize the $\FPeel()$ function in \cref{sec:peeling} and \cref{sec:online-framework}.
Other steps in this algorithm can be parallelized trivially---a parallel partition can generate $\alivevertices$ in \cref{line:filter_out_alive} and $\frontier$ in \cref{line:generate_frontier} using $O(n')$ work and $O(\log n')$ span where $n'$ is the input array size.
}


Note that the two functions on \cref{line:extract} and \cref{line:pack} can be implemented in $O(|\alivevertices|)$ work by the standard parallel \pack{} function introduced in \cref{sec:prelim}. Hence, the key component in this framework is an efficient \FPeel{} function. 
In the following, we will introduce two main peeling approaches for in existing implementations, 
how they can be analyzed using our framework, and
their advantages and drawbacks to process certain types of graphs.

\subsection{Parallel Peeling Process in Existing Work}\label{sec:peeling}
Under the proposed framework, multiple strategies for solving the $\FPeel$ function on~\cref{line:process_bucket} in parallel can be applied. 
We first review the two most relevant strategies in the literature. 

\myparagraph{The Offline Strategy in \Julienne{}.} 
\Julienne{}~\cite{dhulipala2017} employs a batch-synchronous strategy for the $\FPeel{}$ function (\cref{algo:peel_offline}). 
In each \subround, it gathers all vertices that require degree changes in a list $L$, 
which concatenates the neighbor lists for all vertices in $\frontier$. 
Each appearance of a vertex $u$ in $L$ means to decrement the \induceddegree{} $\degreestar[u]$ by 1. 
Hence, a \mf{Histogram} algorithm is used to count the number of appearances for each vertex in the list,
which can be performed by a parallel semisort~\cite{gu2015top,dong2023high} with $O(n)$ work with high probability. 
The \induceddegree{} for each vertex in $L$ will be decremented in a batch accordingly, and the next frontier can be computed as
all vertices that have degree drop to $k$ or lower by a parallel pack. 
We refer to this approach as an \defn{offline} approach. 
Since each vertex and edge is processed exactly once in the peeling process, 
the total work for this step is proportional to the total neighborhood size of the vertices in the frontier.
Based on \cref{thm:work}, the \kcore implementation in \Julienne{} has $O(n+m)$ work.

\begin{algorithm}[t]
  \small
    \caption{Offline peeling process}
    \label{algo:peel_offline}
    \SetKwFor{parFor}{parallel\_for}{do}{endfor}   
    \SetKwProg{myproc}{procedure}{}{}
    \myfunc{\FPeel{$(\frontier, k)$}}{
    $L \gets $ \text{the list of vertices $u$, s.t. $(u,v)\in E, v\in \frontier$; duplicates are kept.} \\
    $H \gets \mf{Histogram}(L)$ \label{func:histogram}\tcp*[f]{Count the frequency for each $u$ in $L$}\\

    \parFor(\tcp*[f]{for each $u$ with frequency $f_u$}) {$(\vname{u}, f_u) \in H$ }{

      \customIf{$\degreestar[u] > k$}{
          $\degreestar[u] \gets \degreestar[u] - f_u$\label{line:offline:decdeg}
        }
      }
    $\nextfrontier\gets \{u\in L, \degreestar[u]$ dropped to $k$ or lower by \cref{line:offline:decdeg}  $\}$ \\
    \Return{$\nextfrontier$}
    }
    \end{algorithm}

The actual implementation of \Julienne{} is more complicated than our framework. 
Our new algorithm with offline peeling is much simpler than both their theoretical algorithm in the paper and their implementation. 
Our result indicates that achieving work-efficiency does not require the complicated bucketing structure. 
We note that adding a bucketing structure can likely improve the performance in practice, which we discuss in \cref{sec:bucketing-existing}. 

While \Julienne{} achieves efficient work, the parallelism of \Julienne{} can be bottlenecked the scheduling overhead caused by the offline algorithm.
Note that each \subround{} requires to distribute work to all threads and synchronize them at the end. 
On sparse graphs, this scheduling overhead can dominate the actual computation cost, 
making \Julienne{} slower than a sequential algorithm in certain cases. 
In \cref{sec:vgc} we discuss more details and provide our solution to overcome this challenge. 

\myparagraph{The Online Strategy in \ParK and \PKC.} 
Both \Park{}~\cite{dasari2014park} and \pkc{}~\cite{kabir2017parallel} use an asynchronous peeling algorithm (\cref{algo:peel_online}), which
removes vertices in the frontier in parallel, 
and directly decrements the \induceddegree{s} of their neighbors using the \atomdec{} operation. 
We refer to this approach as an \defn{online} approach, 
since when we process a vertex $v$ in the frontier, the \induceddegree{s} $\degreestar[\cdot]$ of its neighbors are updated immediately. 
When $\degreestar[u]$ is decremented from $k+1$ to $k$, 
$u$ will be put in the next frontier by atomically appending it to an array $\nextfrontier$. 
Assuming constant work for each \atomdec{} operation and appending each element to $\nextfrontier$, 
each peeling \subround has cost proportional to the total neighborhood size of the vertices in the frontier. 

Unfortunately, neither of the two algorithms maintains the \alive{} set during the algorithm, 
and simply use the original vertex set $V$ to generate each frontier. Therefore, the two algorithms both have $O(m+\maxcoreness n)$ work. 
However, introducing the \alive{} set in the algorithm will directly lead to work-efficiency.

The computation in the online approach is simpler than the offline approach, and does not require strong synchronization between subrounds.   
Certain optimizations can be used to alleviate the scheduling overhead (e.g., \PKC{}~\cite{kabir2017parallel} and our new solution in \cref{sec:vgc}). 
The major performance bottleneck of \park{} and \pkc{} comes from the potential contention caused by the atomic operations. 
For a high-degree vertex, many of its neighbors can decrement its \induceddegree{} concurrently. 
As we can observe from the results in~\cref{table:fulltable}, \park{} and \pkc{} both suffer from poor performance on power-law graphs such as social and web graphs. 
Even a small fraction of high-degree vertices may result in high contention that degrade the performance. 
In addition, the data structure $\nextfrontier$ to generate the next frontier may also suffer from contention when many vertices are appended concurrently. 
In \cref{sec:sampling}, we propose our solution to overcome this challenge.

\begin{algorithm}[t]
  \small
    \caption{Online peeling process}
    \label{algo:peel_online}
    \SetKwFor{parForEach}{parallel\_foreach}{do}{endfor}

    \SetKwProg{myfunc}{Function}{}{}
%
%
%
%

  \myfunc{$\FPeel (\frontier,k)$\label{line:peel}}{ 
    Initialize $\nextfrontier\gets \emptyset$ \tcp*[f]{Buffers the next frontier}\\
    \parForEach{$v\in \frontier$} {
        \parForEach{$u\in N(v)$} {
          $\delta\gets\atomdec(\degreestar[u])$ \label{line:atomic-dec-degree}\tcp*[f]{decrement atomically}\\
          \tcp{The last decrement adds $u$ to the next frontier}
          \customIf{$\delta=k+1$} { Add $u$ to $\nextfrontier$\label{line:update_frontier}} 
        }
    }
    \Return $\nextfrontier$\tcp*[f]{Returns the next frontier}
  }
\end{algorithm}

\revise{
    \myparagraph{Span Analysis.} 
    To analyze the span of the algorithm, \Julienne{}~\cite{dhulipala2017} defined a parameter $\rho$ to represent the number of peeling subrounds, 
    referred to as the \defn{peeling complexity}, and showed that the span of their algorithm is $\tilde{O}(\rho)$ \whp{}.
    For the framework in \cref{algo:framework} with both online and offline peeling algorithms, 
    the same $\tilde{O}(\rho)$ span bound holds (\whp{} for the offline version) following the same analysis in \Julienne{}.         
    However, note that each of the $\rho$ subrounds requires a parallel-for loop, indicating a global synchronization among threads. Therefore, the \emph{burdened span} (defined in \cref{sec:prelim}) is $\tilde{O}(\rho\omega)$. 
    Given $\omega$ as a large constant, when $\rho$ is also large, the parallelism in the algorithm can be limited. 
    In \cref{sec:vgc}, we discuss our new techniques that improve the burdened span, thus parallelism. 
}

\section{Our New Techniques}\label{sec:online-framework}
In this section, we introduce our new techniques to overcome challenges in existing solutions. 
Our algorithm is also based on the online algorithm shown in \cref{algo:peel_online}. 
We use the parallel hash bag introduced in \cref{sec:prelim} 
to maintain the frontiers in the algorithm. 

As discussed, there exist two main challenges in existing solutions that limits parallelism on different types of graphs.
The first occurs on graphs with high-degree vertices, 
where the atomic operations in the online algorithm cause high contention.
The second occurs on sparse graphs with low-degree vertices, where synchronization between subrounds causes high scheduling overhead. 
In this section, we propose our solutions to these two challenges, including a \emph{sampling scheme} to reduce high contention on concurrent atomic operations,
and a \emph{vertical granularity control} approach that uses a local search to reduce the number of subrounds. 

\subsection{Reducing Contention Using Sampling}\label{sec:sampling}

As mentioned, one challenge in the online peeling algorithm is the high contention 
in decrementing the \induceddegree{} of each vertex concurrently, especially for high-degree vertices.
To overcome this challenge, we propose a sampling scheme. 
In particular, when the degree of a vertex $v$ is over a certain threshold, 
we will turn it on \emph{sample mode}. 
When decrementing the \induceddegree{} of a vertex in the sample mode,
instead of atomically decrementing $\degreestar[v]$, 
we will take a sample for it with a certain probability, which we call the \emph{sample rate}. 
Based on the number of samples and the sample rate, we can estimate the expectation of the current \induceddegree{}.
When the expectation is far above $k$, it is unlikely that $\degreestar[v]$ drops to $k$,
and thus we do not need to know its exact \induceddegree{} in the current round. 
In other words, for a high-degree vertex $v$, we do not update $\degreestar[v]$ explicitly before we collect sufficient samples, 
thus avoiding updating $\degreestar[v]$ frequently. 

To do this, we maintain a structure \emph{sampler} for each vertex $v$ (defined in \cref{algo:sample_func}), recording whether $v$ is in the sample mode, its sample rate, and the number of samples taken so far. 
Based on the sampler of a vertex $v$, we can estimate the expectation of the true \induceddegree{} of $v$,
as well as the probability that the true \induceddegree{} is lower than $k$. 
If the true \induceddegree{} is likely to be lower than $k$, 
we will resample $v$ to either use a higher sample rate,
or terminate the sample mode. 

While the idea is simple, the intricate part is to ensure that at peeling round $k$, all vertices with coreness $k$ must not be in sample mode and have their the true \induceddegree{s} computed. 
The challenge is then to control the error probability for any vertex to miss its peeling round. 
Next, we present the details in our sampling algorithm, such that it gives correct answers with high probability. 

\subsubsection{The Framework with Sampling} 
We present our framework with sampling in \cref{algo:sampling}. 
We first introduce this framework at a high level, and then elaborate on each function in \cref{algo:sample_func}. 
The framework roughly follows \cref{algo:framework} with a few changes due to sampling. 
First of all, we start with initializing the sampler of $v$ by $\FInitSampler{}(v,0)$, 
which determines whether $v$ should be sampled when $k=0$, and if so, what its sample rate should be.

\begin{algorithm}[t]
\small
  \caption{Our algorithm framework with sampling \label{algo:sampling}}
  \DontPrintSemicolon
  \textbf{Input:} Graph $G = (V, E)$. \hfill \textbf{Output:} $\truecoreness[\cdot]$: Coreness of each vertex
  \SetKwFor{parFor}{parallel\_for}{do}{endfor}
  \SetKwInOut{Maintains}{Maintains}
  \SetKwProg{myfunc}{Function}{}{}
  \SetKwFor{parForEach}{parallel\_foreach}{do}{endfor}
  \SetKwInOut{Maintains}{}
  Initialize $\alivevertices\gets V$; $k\gets 0$; $\degreestar[v]\gets \degree(v)$ for all $v\in V$ \label{line:set_sampler_frame}\\
    \lparForEach(\tcp*[f]{Initialize $\sampler[v]$}){$v\in V$}{\FInitSampler{$(v, 0)$}}
    \While{$\alivevertices\ne \emptyset$} {
      $\frontier \gets \{v~|~v\in\alivevertices,\degreestar[v]=k\}$\\
      \parForEach{$v\in V: v$ is in sample mode\label{line:framework:security1}}{
      \lIf{$\FError(v,k)=\false$\label{line:framework:security2}}
      {$\FSetSampler(v,k,\frontier)$\label{line:framework:security3}} 
      }
      \While{$\frontier\ne \emptyset$}{
        \lparForEach{$v\in \frontier$}{$\truecoreness[v]\gets k$}
        $\langle \frontier, \countingbag\rangle \gets \FPeel (\frontier,k)$ \label{line:sample_peel}\tcp*[f]{$\countingbag$: vertices that require to reset samplers}\\
        \lparForEach{$v\in \countingbag$}{\FSetSampler{$(v, k,\frontier)$}\label{line:count_vertex_frame}} 
      }
  
      $\alivevertices \gets \{v~|~v\in\alivevertices,\degreestar[v]>k\}$\\ 
      $k \gets k + 1$\\  
    }
    \Return $\truecoreness[\cdot]$
  \end{algorithm}

In general, two conditions may trigger updating the sampler of a vertex $v$. 
The first case is when $k$ approaches the true \induceddegree{} of $v$, 
and thus we need to count $\degreestar[v]$ more accurately. 
In particular, 
when a peeling round starts, 
we validate all vertices in the sample mode remain safe to be in the sample mode until the next round (\cref{line:framework:security1,line:framework:security2,line:framework:security3}). 
If the validation fails, we resample $v$. 


Another case to ressample $v$ is when sufficient samples have been collected, and thus we know the \induceddegree{} of $v$ has dropped significantly. 
This happens in the $\FPeel$ function. 
Therefore, \FPeel{} also identifies a set of vertices $\countingbag$ that have collected sufficient samples and requires resampling. 
After the peeling process, we resample all vertices in $\countingbag$ (\cref{line:count_vertex_frame}).

In the following, we explain the details of the \FPeel{} function under the sampling scheme, as well as the helper functions \FInitSampler{} and \FSetSampler{} to handle the samplers of each vertex. 

\begin{algorithm}[t]
\small
    \caption{
    Functions used in our algorithm with sampling
    } 
    \label{algo:sample_func}
    
    \DontPrintSemicolon

%


\let\oldnl\nl
\newcommand{\nonl}{\renewcommand{\nl}{\let\nl\oldnl}}

\SetKwFor{parFor}{parallel\_for}{do}{endfor}
\SetKwFor{parForEach}{parallel\_foreach}{do}{endfor}
\SetKwInOut{Parameters}{Parameters}
\Parameters{
$\reducerate$: when $\degreestar[v]$ decrement to a factor of $\reducerate$, we resample $v$\\
$\mu=4c\ln n$: expected number of hits each sampler, $c>2$
  }
  \SetKwProg{mystruc}{struct}{}{}
  \nonl \mystruc(\tcp*[f]{For each $v\in V$, maintain a \emph{sampler} structure $\sampler[v]$}){sampler} {
   \nonl $\vname{\mode{}}$\,: boolean flag indicating whether $v$ is in sample mode\\
   \nonl $\vname{\rate{}}$\,: the sample rate for $v$\\
   \nonl $\vname{\cnt{}}$\,: the number of hits in the sampling process\\
  } 
\smallskip
  \myfunc(\tcp*[f]{peeling process with sampling}){$\FPeel (\frontier,k)$}{ 
    $\nextfrontier\gets \emptyset$; $\countingbag\gets \emptyset$\tcp*[f]{$\countingbag$: vertices to recount their \induceddegree{s}}\\
    \parForEach{$v\in \frontier$} {
        \parForEach{$u\in N(v)$} {
          \If(){$\sampler[u].\mode$}{
            $\delta\gets \atominc(\sampler[u].\cnt)$ with probability $\sampler[u].\rate$ \label{line:sample_vertex_success}\\
            \tcp{If sufficient samples are collected, add $u$ to $\countingbag$}
            \lIf{
                $\delta=\exphits-1$ \label{line:sample_hit_thres}
            }{
                Add $u$ to $\countingbag$  \label{line:sample_disable_and_count} 
            }            
          } 
          \Else {
            $\delta\gets\atomdec(\degreestar[u])$\\
            \lIf{$\delta=k+1$} {Add $u$ to $\nextfrontier$} 
          }            
        }
    }
    \Return $\langle \nextfrontier, \countingbag\rangle$
  }
  
\myfunc{\FInitSampler{$(v, k)$}}{
    \label{algo:init_sample_func}  
    \tcp{If the expected \induceddegree{} of $v$ is still large and far from $k$ 
    even after decrementing to a factor of $\reducerate$, then $v$ can be sampled safely. 
    }
    \If{
        ($\degreestar[v] \cdot \reducerate{} > k) \wedge (\degreestar[v]>$ threshold\,$)$  \label{line:sample_cond}
    }{
        $\sampler[v].\mode \gets \true$ \label{line:set_sample_mode}\\
        \tcp{Set the sample rate. This formula is explained in \cref{sec:sampling}.}  
        $\sampler[v].\rate\gets \exphits / ((1 - \reducerate) \cdot \degreestar[v])$ \label{line:set_sample_rate} \\
        $\sampler[v].\cnt\gets 0$
    }
    \lElse{
        $\sampler[v].\mode \gets \false$
    }
}

\myfunc{\FSetSampler{$(v, k, \frontier)$}}{
    \label{algo:set_sample_func}  
    $\degreestar[v]\gets$ the number of \alive{} vertices in $\nei(v)$\\
    \lIf{$\degreestar[v]\le k$}{Add $v$ to $\frontier$}
    $\FInitSampler(v,k)$
}

\myfunc(\tcp*[f]{Explained in \cref{sec:calc_error}}){\FError{$(v, k)$}\label{line:check-resample}}{
    \Return {\upshape $(\degreestar[v]\cdot r>k)\wedge(\sampler[v].\cnt<\sampler[v].\rate\cdot(\degreestar[v] - k)/4)$}
}

\end{algorithm}

\subsubsection{Details about the Sampling Scheme} 
\label{sec:sampling:details}

As mentioned, each vertex $v$ maintains a \emph{sampler} structure including three fields: a boolean flag $\mode$ indicates whether $v$ is in the sample mode, a float $\rate$ as the sample rate, and an integer $\cnt$ as the number of samples taken by $v$ so far. 
In general, with sampling rate $p$ and $s$ samples taken, we expect the \induceddegree{} to reduce by $s/p$. 
However, to enable a high probability guarantee for the estimation $s/p$, the number of samples need to be $\Omega(\log n)$ (see \cref{sec:calc_error}). 
In our algorithm, we use a parameter $\exphits=\Theta(\log n)$ to denote the desired number of samples we need to take before we ressample $v$. 
Namely, we wish to take at least $\exphits$ samples to ensure high confidence for the estimation of the \induceddegree{} for $v$. 
When the \induceddegree{} of $v$ has decremented by a significant factor, 
the sample rate has to be adjusted (increased) accordingly. 
In our case, when the \induceddegree{} drops to a factor of $\reducerate$, we resample $v$. 
In practice, we use $\reducerate=10\%$. 
In other words, we use the current sampler to estimate the \induceddegree{} of $v$, 
until we expect the true $\degreestar[v]$ drops to $r\cdot \degreestar[v]$, at which point we reset the sampler of $v$. 

We start with presenting our \FInitSampler{} function, which initializes the sample parameters for a vertex $v$. 
Based on the discussion above, we hope that when $\exphits$ samples have been taken, 
we expect the \induceddegree{} of $v$ drops from $\degreestar[v]$ to $\reducerate\cdot \degreestar[v]$. Therefore, the sample rate should be set as $\exphits/((1-\reducerate)\cdot\degreestar[v])$ (\cref{line:set_sample_rate}). 
Accordingly, to determine whether a vertex is safe in sample mode, $\reducerate\cdot\degreestar[v]$ must still be large enough.
In our case, we require it to be larger than a preset threshold, as well as $k$. 
The latter condition is because when the expected \induceddegree{} approaches $k$, the probability that the true \induceddegree{} is smaller than $k$ will increase dramatically. 
If $\reducerate\cdot\degreestar[v]$ is larger than both the sampling threshold and $k$, 
we turn on the sample mode for $v$, and set the parameters accordingly. 

We then present our \FPeel{} function with sampling in \cref{algo:sample_func}. 
This function follows the online peeling process in \cref{algo:peel_online}. 
The only difference occurs at \cref{line:sample_vertex_success,line:sample_hit_thres} in \cref{algo:sample_func}, when peeling $v$ and decrementing the \induceddegree{} of its neighbor $u$. 
If $u$ is in the sample mode, this indicates that $u$'s current \induceddegree{} is far above $k$. 
Therefore, instead of decrementing $\degreestar[u]$, we increment the number of samples of $u$, stored in $\sampler[u].\cnt$, with probability $\sampler[u].\rate$. 
This increment is also performed atomically, as multiple threads may be accessing $u$ concurrently. 
If the atomic increment causes $\sampler[u].\cnt$ to reach the desired number of samples $\exphits$, we add $u$ to $\countingbag$, 
which buffers all vertices that requires resampling. 

To resample a vertex $v$, we use function $\FSetSampler(v,k,\frontier)$. 
It begins by counting the actual number of \alive{} vertices of $v$, giving the true value of $\degreestar[v]$.
If $\degreestar[v]$ reaches $k$ or lower, the vertex will be added to the current frontier.
Then we call $\FInitSampler$ to reset the parameters for the sampler, based on the new value of $\degreestar[v]$. 

We note that due to sampling, it is possible that a vertex $v$ is still in the sample mode when the peeling round $k=\truecoreness[v]$. 
If so, we are unable to peel $v$ correctly because we do not know its true \induceddegree{} at that time. 
However, we will show that 1) it happens with low probability, and 2) if this happens, we can always detect it, 
and can restart with stronger sampling parameters. 
In \cref{sec:calc_error}, we introduce our approach to handle such potential errors. 

\subsubsection{Correctness Analysis} \label{sec:calc_error}
We now analyze the correctness of our sampling scheme, and show the error probability is low.
The only case for an error is when vertex $v$ is in the sample mode, but the true \induceddegree{} of $v$ is less than $k$ at the beginning of round $k$. 
We will show that using our $\FError()$ check, the error probability is very low, based on our choices of parameters ($\sampler[v].\rate$ and $\sampler[v].\cnt$).

\begin{lemma}\label{lem:coin-toss}
  Assume we toss $t$ coins each with $p$ probability to be a head, and obtain $s$ heads. 
  If $tp\ge 4c\ln n$, then $s\ge tp/4$ \whp{}.  
\end{lemma}
\begin{proof}
  This can be shown by using the lower part of the of the multiplicative form of the Chernoff bound.
  Let $s$ be the number of heads seen.
  In this case, $\mu=tp$, and $\delta=1-s/tp$.
  Hence we have:
  \begin{align*}
        \Pr\left[s<{tp\over 4}\right] &\le\exp\left(\delta^2\mu\over 2\right)=\exp\left(s-{s^2\over 2tp}-{tp\over2}\right)\\
        &<\exp\left(s-{tp\over 2}\right)<\exp\left(-{tp\over 4}\right)=n^{-{tp\over4\ln{n}}}=n^{-c}.
  \end{align*} 
  This proves the lemma.
\end{proof}

With \cref{lem:coin-toss}, we now prove that \cref{algo:sampling} is correct \whp{}.

\revise{
  \begin{theorem}\label{thm:sampling}
    For any constant $c\ge 1$, using $\mu = 4(c+2)\ln(n)$, \cref{algo:sampling} is correct with probability $1-n^{-c}$. 
  \end{theorem}
  \begin{proof}

  For a vertex $v$ in sample mode, we use $d^*$ to denote its true \induceddegree{}, and $\degreestar[v]$ is the \induceddegree{} at the point when we last call $\FInitSampler$ on it (either at the beginning of the algorithm or the last time of resampling). 
  In \cref{lem:coin-toss},
  each increment of sample count on \cref{line:sample_vertex_success} is a coin toss with probability $p=\sampler[v].\rate$. 
  If $d^*$ drops to $k$ or lower, at least $t=\degreestar[v] - k$ edges have been removed, corresponding to $t$ coin tosses, 
  and the expected number of sample count is $tp=\sampler[v].\rate\cdot(\degreestar[v] - k)$. 

  For a vertex $v$ in the sample mode, we first show that if its induced degree $d^*<k$, then \FError{} returns \false{} \whp{}. 
  First of all, if $k\ge r\cdot \degreestar[v]$, the function will fail at the first condition. 
  Otherwise, $k< r\cdot \degreestar[v]$. 
  In this case, $tp=(\degreestar[v] - k)\cdot\sampler[v].\rate$. Plugging in $\sampler[v].\rate=\mu/((1-r)\cdot \degreestar[v])=4(c+2)\ln n / ((1-r)\cdot\degreestar[v])$, we have $tp=4(c+2)\ln n \cdot ((1 - k/\degreestar[v])/(1-r))$. 
  Combining with the assumption that $k< r\cdot \degreestar[v]$, we have
  $tp\ge 4(c+2)\ln n$. 
   
  Based on \cref{lem:coin-toss}, 
  this means that we should have collected $s\ge tp/4\ge(c+2)\ln n$ heads (successful samples) \whp{}. 
  Plugging in $s=\sampler[v].\cnt$, $t=\degreestar[v]-k$ and $p=\sampler[v].\rate$, 
  this means \FError{} will fail at the second condition. 
  Therefore, if $d^*<k$, \FError{} will return \false{} \whp{}.
  At this point, we have shown that the error probability for a vertex $v$ to be in sample mode when $d^*<k$ is smaller than $n^{-(c+2)}$ for any constant $c>0$. 
  
  Note that to make the algorithm correct, all the function calls to \FError{} have to be correct. 
  In \cref{algo:sampling}, $\FError$ can be called for at most $\kmax n\le n^2$ times. 
  By the union bound, the probability that there exists a failed validation is at most $n^{-c}$. 
  Therefore, the algorithm is correct with probability at least $1-n^{-c}$. 
  \end{proof}
  
  
  
  \hide{
    The high probability bound means that the error rate for one invocation of \FError{} is smaller than $n^{-c}$ for any constant $c>0$. 
    In \cref{algo:sampling}, $\FError$ can be called for at most $\kmax n\le n^2$ times. 
    In our algorithm implementation, we set $c=8$ to ensure the correctness of the algorithm.
    The union bound guarantees that the probability of detecting an error for the whole algorithm is smaller than $n^{-8}$.
  }
  
  From \cref{thm:sampling} and the definition of high probability in \cref{sec:prelim}, we have the following corollary. 
  
  \begin{corollary}\label{cor:sampling}
    \cref{algo:sampling} is correct with high probability. 
  \end{corollary}
  \vspace{-0.05in}

Note that error may occur at an unsampled vertex, if one of its neighbors is in sample mode and erred. 
  However, such a case must be caused by an error from a sampled vertex. Thus, the algorithm is correct as long as all sampled vertices are processed correctly. 
}

\subsubsection{Recover from Errors}

Although \cref{thm:sampling} guarantees that \cref{algo:sampling} succeed \whp{}, we still need to detect and correct any possible errors caused by sampling, so that \cref{algo:sampling} is \emph{Las Vegas} instead of \emph{Monte Carlo}.
A possible error happens when a vertex $v$'s induced degree drops below $k$ in the sample mode before round $k$.
This can be detected when $v$ exits the sampling mode, when we count the true value of $\degreestar[v]$.
Note that even if this value is smaller than $k$ in round $k$, it may not be an error since in normal peeling process, 
when peeling vertices in subrounds, some vertices in $k$-core may have $\degreestar[v]$ drops below $k$.
To verify the correctness, our algorithm will further check whether $v$'s \induceddegree{} is at least $k$ in the previous round.
If so, $v$ is still safe and have coreness $k$. 
Otherwise, we can restart the algorithm with a larger value of $\exphits$ or without sampling. 
However, due to the strong theoretical guarantee \cref{thm:sampling} provides, 
we have never encountered restarting in executing our \kcore algorithm. 

\subsubsection{Cost Analysis}\label{sec:sampling:contention}
Sampling does not affect the work-efficiency of our algorithm. 
given our setup of our sampling scheme, when a vertex $v$ exits the sampling mode with sufficient samples, $\degreestar[v]$ is reduced by a constant fraction \whp{}.
Therefore, the total cost to recount the true \induceddegree{} of $v$ is $O(\degree(v))$, which add up to at most $O(m)$. 
The cost validation step is proportional to the number of vertices in the sample mode, and thus is bounded by the active set size. Therefore, the proof in \cref{thm:work} still holds. 

Sampling can also reduce contention.
\revise{
  As defined in \cref{sec:prelim},
  The contention is the number of possible concurrent operations.   
  If a vertex $v$ is in the sample mode, the concurrent updates will be performed on $\sampler[v].\cnt$. 
  For all its $\degreestar[v]$ remaining neighbors, each of them will increment $\sampler[v].\cnt$ with probability $\sampler[v].\rate=\exphits/((1-r)\cdot \degreestar[v])$. Therefore, the contention on $\sampler[v].\cnt=\exphits/(1-r)=O((\log n)/(1-r))$. 
  If a vertex is not in the sample mode (or it has terminated from sample mode), base on the if-condition on \cref{line:sample_cond}, the \induceddegree{} is at most $O(k/r+\mathit{threshold})$. 
  Considering both $r$ and $\mathit{threshold}$ are preset as constants, combining both cases, the contention to handle a specific vertex $v$ is $O(\truecoreness[v]+\log n)$. 
  For high-degree vertices, this is much smaller than $O(d(v))$, which is the number of concurrent updates to $\degreestar[v]$ without sampling.  
}

\subsection{Vertical Granularity Control}\label{sec:vgc}

\begin{figure}[t]
  \centering
  \includegraphics[width=0.85\columnwidth]{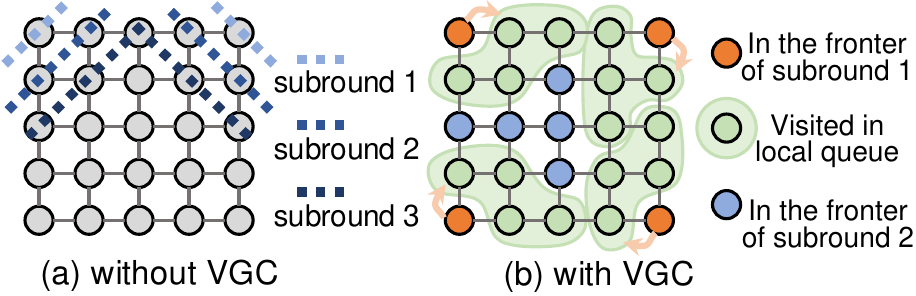}
  \caption{The peeling process on a grid with and without using VGC.
  In this example, the queue size is 4.
  Note that the execution of VCG is not deterministic, and (b) shows a possible execution.}\label{fig:localsearch}
\end{figure} 

Our second approach tackles the significant scheduling costs that occur in sparse graphs. 
In the framework given in \cref{algo:framework}, each \subround{} must wait for the completion of the previous one. 
Between these subrounds, the scheduler distributes tasks to all threads and synchronizes them at the end, which can incur substantial scheduling overhead. 
This overhead is particularly pronounced in sparse graphs, where vertices typically have low degrees. 
The computation for processing a low-degree vertex $v$ is small, meaning that the overhead associated with creating and synchronizing the thread for $v$ can outweigh the computational costs. 
As a result, this overhead diminishes the benefits of parallelism.

Our approach to overcome this challenge is inspired by a recent technique~\cite{wang2023parallel,dong2024pasgal} known as \emph{vertical granularity control (VGC)}. 
Similar to traditional granularity control in parallel programming, VGC focuses on increasing the size of base case tasks in graph processing to mitigate scheduling overhead. 
In our \kcore{} algorithm, we aim to reduce the overall number of subrounds and eliminate small parallel tasks. 
To achieve this, we incorporate VGC into our online peeling process in \cref{algo:peel_online}.
When peeling a low-degree vertex $v$, we place all its active neighbors in a FIFO queue, referred to as the \emph{local queue} of~$v$, 
and process all vertices in the local queue sequentially. 
When we decrementing the \induceddegree{} of a neighbor $u$,  
if $\degreestar[u]$ drops to $k$ (\cref{line:update_frontier} in \cref{algo:peel_online}),
instead of adding $u$ to $\nextfrontier$, we add $u$ to the local queue. 
This allows $u$ to be processed in the same subround as $v$, rather than waiting for the next subround.
We refer to this process as a \emph{local search} at $v$. 

\revise{There are multiple ways to control the granularity in VGC, 
such as controlling the local queue size, 
or the number of touched vertices, to be a fixed number. 
In our code, we simply fix the local queue size as $128$. 
Once the local queue is full, even if $\degreestar[u]$ drops to $k$, 
we still add $u$ to the next frontier as normal. }
An illustration of VGC on a grid is given in \cref{fig:localsearch}. 
\revise{Limiting the work of each local search caps the maximum work each thread receives, 
thereby ensuring better load balancing. }
Empirically, processing hundreds of vertices is sufficient to hide scheduling overhead~\cite{wang2023parallel}. 
In our experiments, performance remains relatively stable across queue sizes ranging from hundreds to thousands.

VGC does not change the work-efficiency of the algorithm, as each vertex is still processed exactly once, 
either from the local queue or from the frontier. 
In practice, VGC improves the performance for sparse graphs by 1.72--31.2 times. 

Some previous work also tried to reduce the number of subrounds. 
In \PKC{}~\cite{kabir2017parallel}, each processor keeps a subset of $\frontier$ in a local buffer, and performs a sequential peeling process until the buffer is empty, giving exactly one \subround per round. 
However, it may cause severe load imbalance---if one vertex triggers a chain reaction in peeling, 
the corresponding processor may perform significantly more work than others. 
In contrast, our algorithm parallelizes all vertices and use a work-stealing scheduler to enable dynamic load-balancing, 
and use VGC on top of it to hide scheduling overhead. 
In summary, VGC improves performance by both hiding synchronization cost between subrounds and achieving good load balancing.

\revise{
    \myparagraph{Burdened Span Analysis.} 
    To analyze the improvement of VGC, we use the \emph{burdened span} defined in \cref{sec:prelim}, which charges a factor of $\omega$ to each fork/join operation in the span to reflect scheduling overheads. 
    Recall that the burdened span of the framework in \cref{algo:framework} (e.g., the offline algorithm in \Julienne{}) is $\tilde{O}(\omega \rho)$,
    where $\rho$ is the number of subrounds in the algorithm. 
    Let $\rho'$ be number of subrounds in our algorithm with VGC, and $L$ the total work performed by each local search.
    In this case, our algorithm with VGC has a span of $\tilde{O}(\rho' (\omega + L))$. 
    As mentioned, we can control $L$ (the granularity of VGC) in various ways, and based on the theory, the primary goal
    of VGC is to control $L$ asymptotically lower than $\omega$, where $\omega$ is on the order of magnitude of $10^4$~\cite{he2010cilkview}. 
    Since $\rho' \leq \rho$, the burdened span of our algorithm with VGC is always no worse than the version without VGC. 
    In most of the cases, $\rho' < \rho$, and thus the burdened span can be improved by VGC. 
    In \cref{fig:queue-rho}, we empirically showed that the number of subrounds after VGC (i.e., $\rho'$) can be 5--40 times smaller than $\rho$ on the tested graphs. 
    This justifies why our solution with VGC can achieve better parallelism than \julienne{}. 
}

\section{Hierarchical Bucketing Structure}\label{sec:bucketing}

\cref{thm:work} shows that explicitly checking the active vertices in $\alivevertices$ (\cref{line:filter_out_alive}) in each round is asymptotically optimal.
However, this simple solution can be slow in practice. 
To optimize the performance, some existing designs, 
such as \Julienne{}, choose to use a bucketing structure (defined in \cref{sec:prelim}) to maintain the active set $\alivevertices$. 
A bucketing structure usually maintains a (partial) mapping from each value $d$ to all vertices in $\alivevertices$ with (induced) degree $d$. 
In this section, we introduce our new design for the bucketing structure.  

We proposed the \emph{hierarchical bucketing structure} (\HBS), and show
how it improves our \kcore{} algorithm. 
We note that this data structure may also be of independent interest, 
since it provides the interface a special parallel priority queue with integer keys, which is useful in many applications~\cite{li2013parallel, shi2021parallel, shi2023theoretically,dhulipala2017,gbbs2021}.


\subsection{Interface and Related Work}\label{sec:bucketing-existing}


As mentioned, while refining the active set in each round does not increase the asymptotic cost, it can affect performance due to extra computations and memory accesses. 
To reduce the overhead, \Julienne{} implements a bucketing structure, 
which maintains a collection of elements (vertices), each of which has a unique identifier (vertex ID) and an integer key (induced degree). 
The structure supports the following three functions (we note that there are more functions in the interface of a bucket structure~\cite{dhulipala2017}; For simplicity, we only list the functions used in \kcore):

\begin{itemize}
    \item \textbf{\FBuildBuckets{$R$, $A$}}: initialize the bucketing structure with key range 0 to $R$, and insert each element $a\in A$ into it.
    \item \textbf{\FGetNextBucket~~$\mapsto \frontier$}: return all elements with the smallest key in the bucketing structure.
    \item \textbf{\FUpdate{$a$}}: update $a$ in the structure with its new key.
\end{itemize}

\Julienne{} maintains a bucketing structure with $b$ buckets. Every $b$ rounds, it generates all frontiers for the next $b$ rounds by $\FBuildBuckets{b,\alivevertices}$,
which extract vertices with induced degree $k+i$ to bucket $i$ ($k$ is the current peeling round). 
Vertices with \induceddegree{s} more than $k+16$ remains in $\alivevertices$, which they call a \emph{overflow bucket}. 
Using an efficient algorithm, the next $b$ frontiers can be generated by one pass of memory access to $\alivevertices$, leading to better performance. 
This strategy reduces the number of accesses to $\alivevertices$ by a factor of $b$. 
A vertex $v$ will be accessed by $\FBuildBuckets{}$ until it is extracted from $\alivevertices$, which means $O(\deg(v)/b)$ times of accesses.
However, when the induced degree of $v$ is decremented, 
we also need to move $v$ to the new bucket by $\FUpdate{v}$.
In the worst case, a vertex may be moved $b-1$ times across the buckets.
Therefore, the total cost to process $v$ in the bucketing structure is $O(\degree(v)/b+b)$, adding up to $O(m/b+nb)$ for all vertices. 
This function has its minimum value at $b=O(\sqrt{\dbar})$, where $\dbar=\Theta(m/n)$ is the average degree. 
Our framework can be viewed as a special case where $b=1$, which only extracts the next frontier (bucket). 


In practice, \Julienne{} uses $b=16$, which is a reasonable trade-off.
Indeed, in our experiments, using $b=16$ provides better performance than $b=1$ on most graphs with a large $\dbar$. 
However, we still observe some challenges with this solution. 
First, on most graphs with a small $\dbar$, using 16 buckets may cause 20--70\% overhead than using a single bucket.
Second, on graphs with very large $\dbar$, the cost of $O(\sqrt{m/n})$ per vertex may still be significant. 
Next, we propose our new design to overcome these issues. 

\begin{figure}[t]
  \centering
  \includegraphics[width=.65\columnwidth]{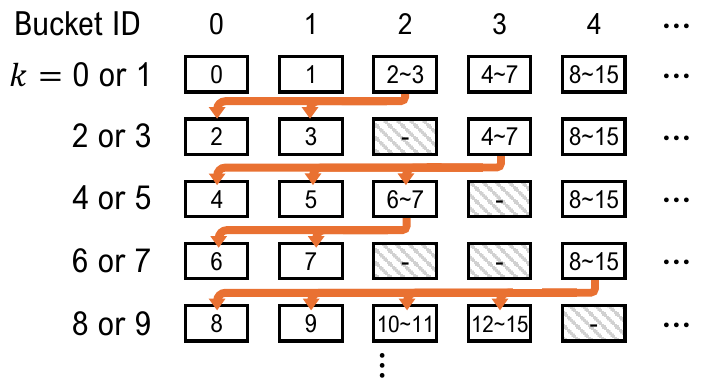}
  \caption{\textbf{The execution of the hierarchical bucketing structure for the first 10 rounds of execution.}  The number in each box indicates the key (\induceddegree) range of the associate bucket. \revise{
    The first row shows that a vertex with degree $d$ is initially inserted to bucket $\lceil\log_2 (d+1)\rceil$.
    When $k=0$ or $1$, vertices with degree 0 or 1 are directly extracted from buckets 0 and 1. 
    We then redistribute vertices in bucket 2 (with degrees 2 or 3) to buckets 0 and 1 (shown in the second row), 
    such that they can be directly identified when $k=2$ or 3. 
    Similarly, after that, we redistributed vertices in bucket 3 (with degrees 4 to 7) to the first three buckets. 
    Vertices with degree 4 and 5 are moved to bucket 0 and 1, respectively, and vertices with degree 6 and 7 are moved to bucket 2, so on so forth. 
}
}
  \label{fig:hier_bucketing}
\end{figure}

\subsection{Hierarchical Bucketing Structure (\HBS)}
We first propose the \emph{hierarchical bucketing structure} (\HBS) to improve the $O(\sqrt{m/n})$ cost per vertex. 
Let $d_{\max}$ be the maximum degree in the graph. 
An \HBS maintains $1+\lceil\log_2 (d_{\max}+1)\rceil$ buckets. 
Each bucket stores vertices within a range of (induced) degrees of 1, 1, 2, 4, ..., growing exponentially. 
\cref{fig:hier_bucketing} illustrates an \HBS and the (induced) degree range that each bucket maintains in the first ten rounds of execution.
When the induced degree of a vertex $v$ drops across a boundary, $v$ is moved to the appropriate new bucket. 

To use an \HBS in \cref{algo:framework}, we will call \FBuildBuckets{$d_{\max}$, $V$} at the beginning of the algorithm, use \FGetNextBucket{} to generate the frontier at \cref{line:generate_frontier}, and call \FUpdate{$v$} when we decrement $v$'s induced degree. 
During the execution, we keep a counter~$\dmin$ indicating the current minimum key in a \HBS{}, which is also the current round $k$.
Finding the bucket ID of an element is based on the most significant differing bit between its key and $k$.

Next, we show how these functions are implemented efficiently on \HBS{}.
We will maintain each of the $O(\log d_{\max})$ buckets by a parallel hash bag (see \cref{sec:prelim}) that supports $O(1)$ expected cost for insertion and $O(t)$ work to extract all $t$ elements from the bag.
\FBuildBuckets{} simply inserts all elements to the corresponding bucket in parallel, based on the most significant bits of their key.

When \FGetNextBucket{} is called, 
we find the first non-empty bucket with id $j$.
If it is one of the first two lists, we directly call \bagpack{} to return all keys in it.
Otherwise, 
if $j>2$, we call \bagpack{} to get all elements in this bucket, and redistribute them to the first $j-1$ buckets, 
shown as the arrows in \cref{fig:hier_bucketing}.

Finally, when an \FUpdate{$a$} is called, we check $a$'s new bucket ID and insert it to the corresponding bucket.  
We do not delete $a$ from the original bucket since a hash bag does not support deletion.
In this case, an element may have copies in multiple buckets. Hence, when we call \bagpack{}, 
we also filter out those elements if their keys does not match the bucket ID. 
We note that this does not increase the asymptotic work of \kcore{}---since a vertex $v$ may be have at most $O(\log d(v))$ copies, the total cost here is at most $\sum \log d(v)=O(m)$. 

In our implementation, instead of setting degree ranges for each bucket as 1, 1, 2, 4, 8, ... , 
we set the first eight buckets as single-key bucket. 
In other words, the first 8 buckets each maintain vertices with degree $\dmin,\dmin+1, \ldots, k+7$, 
and the next buckets correspond to the ranges of $[\dmin+8,\dmin+15], [\dmin+16, \dmin+31]$, and so on.
This optimization avoids frequently redistributing vertices in small buckets. 

\myparagraph{Cost Analysis.}
The cost of \HBS{} on a vertex $v$ includes inserting $v$ (in \FUpdate{}) or redistributing $v$ (in \FGetNextBucket{}) to a new bucket. 
In \FBuildBuckets{}, $v$ is placed in the \revise{$\lceil\log_2 (\degree(v)+1))\rceil$-th} bucket, and will only move to buckets with smaller IDs. 
During \FUpdate{} or \FGetNextBucket{}, $v$ can be packed from each bucket for redistribution at most once, and be inserted to any bucket at most once. 
Hence, the total cost to access $v$ in the bucketing structure is $O(\log \degree(v))$. 
Recall that using a fixed number of $b$ buckets leads to $O(\degree(v)/b+b)$ cost for vertex $v$, 
where $\degree(v)/b+b\ge 2\sqrt{\degree(v)}$.
Therefore, our new design provides a better solution. 

\subsection{Our Final Design}

We now propose our final design of \HBS{}.
The cost to handle vertex $v$ in \HBS{} is $O(\log \degree(v))$, 
compared to $O(\sqrt{m/n})$ using a fixed number of buckets, or $O(\degree(v))$ if no bucketing structure is used.
We first note that when the average degree is constant, 
using a bucketing structure offers no benefit, 
as the overhead may introduce a large hidden constant in the complexity.
Therefore, we only use our \HBS{} when the average degree is larger than a constant $\theta$, which is set to 16 in our code.
Note that even if the average degree of the original graph is lower than $\theta$, 
as more low-degree vertices are peeled, the average degree of the graph may become larger. 
Ideally, when the average degree surpasses $\theta$, we should switch to \HBS{}. 
In our code, we simply switch to \HBS{} when a $\theta$-core is reached, 
which is guaranteed to have an average degree of at least $\theta$. 

%

\hide{
\myparagraph{Hybrid Bucketing Structure with Heuristic Threshold}
In practice, we still observed an overhead of hierarchical bucketing structure or
a fixed large number of buckets for large-diameter graphs, which have a small $\maxcoreness$ value in general.
such as road networks and some synthetic graphs.
In fact, it is theoretically most efficient to compute the $k$-core decomposition in a single bucket for large-diameter graphs under our framework
to avoid heavy movements between the buckets of the vertices,
since most of the vertices will be reduced to a smaller degree value in the end of the computation.
To reduce the effect of the two bucketing structure,
it is natural to combine the two bucketing structure in a hybrid way to hide the overhead of the hierarchical bucketing structure.
For any type of graphs, we first compute a small range in $(0, \Delta)$ of degree increment in a single bucket.
If the degree increases over $\Delta$, we switch the buckets into the hierarchical bucketing strategy without any effect to the computing process.
In experimental results shown in ~\ref{para:exp_bucketing}, there is no significant overhead for the hybrid bucketing structure for dense social networks and web networks,
but can accelerate the algorithm on road networks.
}

\hide{
We show the bitwise operation for \FUpdate{$\cdot$} function in \cref{algo:HBS_update}.
Essentially, we use the bitwise operation ``leading zero count" to find the first bit that is different between $d$ and 0, and use this bit to decide the returned key value.
Thus the function is $O(1)$ work.
For the \FGetNextBucket function, we simply add the bucket\_id by 1 if $\frontier$ is empty, and otherwise return the same bucket\_id.

\begin{algorithm}[h]
    \small
    \caption{\FGetBucket $()$ in \HBS}
    \label{algo:getbucket}
    \KwIn{identifier $a$ (vertex) with its key $\degreestar[a]$}
    \KwOut{$a$'s bucket index}
    \SetKwInOut{Maintains}{Maintains}
    \Maintains{
        $base\_k$: the key value of the first bucket\\  
        $\degreestar[a]$: the \induceddegree of vertex $a$\\
    }
    
    \SetKwProg{myproc}{procedure}{}{}
    \myproc{\FGetBucket{$(a)$}}{
        \tcc{The function returns the bucket index of the indentifier $a$, 
        and is used in the \HBS algorithm to determine the bucket index of a vertex.}
        \tcp{Check if the identifier is in the first 2 buckets}
        \If{$\degreestar[a] < base\_k + 2$}{
            \textbf{return} $\degreestar[a] - base\_k$ \\
        }
        \Else{
            \tcc{Calculate the index for \HBS based on bit difference, $\oplus $ is the bitwise XOR operation, $\mf{leading\_zero\_count}$ is the number of leading zeros}
            \textbf{return} $\gets 64 - \mf{leading\_zero\_count}(\degreestar[a]$ $ \oplus $ $base\_k)$ 

        }
    }
\end{algorithm}

    The \HBS supports the workflow of $k$-core decomposition in an continuous manner.
    First, \textbf{BuildBuckets($b_1$, $b_2$)} initializes the buckets with the number of frontier-bucket $b_1$ and range-bucket $b_2$.
    Then, all the vertices $V$ are mapped to the buckets based on their degree values using \textbf{MakeBuckets($V, deg[\cdot]$)}.
    In each round, the vertices in the frontier are processed in the frontier buckets.
    \textbf{GetNextBucket()} returns the next available bucket with the vertices in it,
    which are peeled in the next round.
    When the frontier buckets are empty (\textbf{GetNextBucket()} returns a pair with $bucket\_id id_{next}$  of range-buckets),
    the vertices in the next available range buckets are moved forward to the buckets using \textbf{UpdateBuckets(id\_next)}.
    Once \textbf{GetNextBucket()} returns \textsc{NULL},
    \textbf{MakeBuckets($V', deg[\cdot]$)} is called again to map the remaining vertices that have not been processed yet to the hierarchical buckets for the next round.

\myparagraph{Analysis of the Hierarchical Bucketing Structure}
The work-efficiency of our algorithm does not rely on the number and structures of buckets.
However, in practical scenarios, we observe that performance varies with the number of buckets used for different graphs.
Every vertex $v \in V$ will be removed from the frontier exactly once.
For a vertex $v$ with degree $\degree(v)$, assume that $v$ has a final \induceddegree{} value of $k'$.
Thus, the total number of insertions into hashbags are not fixed for different bucketing numbers.
From the start of the insertion into the buckets, the insertions of a vertex $v$ are a series of footsteps of insertions until $v$ is finally peeled from the frontier. 
For example, assume a vertex $v_e$ is inserted into the bucket $b_3$ with range $4-7$ in the first round, and finally peeled in bucket $b_1$ with degree value $1$,
then $v_e$ will lead 3 insertions in the hashbags, i.e., $b_3$, $b_2$, and $b_1$.
Then there are two possible workload before $v$ is finally removed from the frontier for different bucketing strategies:

\begin{itemize}
    \item When using a hierarchical bucketing structure,
    vertices are inserted into the $\log n$ buckets,
    resulting in an insertion overhead work of $O(\sum_{i \in V} \log d_i)$.
    \item Using a single bucket entails batch assignment of vertices for $k_{max}$ rounds,
    The dominant factor of the work is the assignment of vertices to the buckets.
    where each vertex $v \in V$ is assigned to a bucket exactly once.
    The dominant factor in this case is the work of assignment that is $O(\sum_{i \in V} d_i)$.

\end{itemize}

The overhead of hashbag operations is balanced by the structure of the buckets.
The $k$-core decomposition algorithm in \Julienne~\cite{dhulipala2017} implements the only multiple-bucket algorithm.
However, with a fixed number of buckets specified in advance, their relative performance is suboptimal for graphs with varying structures.
To address this and enhance adaptability to different graphs,
we design a dynamic bucketing strategy that maps vertices into a single bucket or hierarchical buckets based on the number of peeling rounds, $k$.
As $k$ increases,
maintaining a fixed number of buckets causes the overhead of hashbag operations to increase.
Our implementation dynamically adjusts the bucketing structure based on the computational process,
balancing the two workload by reducing the two overhead
without manually setting a fixed number of buckets for different graphs in advance.

In Figure~\ref{fig:hier_bucketing}, we illustrate the first three bucketing mappings with $k$ increasing.

Although the hybrid bucketing structure is a trade-off between the two bucketing strategies for different types of graphs,
there is adversarial cases can be built to incur overhead.
For example, a graph with a $\maxcoreness$ of $k_m = \beta + 1$ will always force our algorithm to establish the hierarchical bucketing structure,
where $\beta$ is the threshold for the triggering of the hierarchical bucketing structure.
In this case, unnecessary overhead will be incurred for the hierarchical bucketing structure,
because the algorithm will finish in another round with a single bucket structure that is more efficient.
As we can see in experimental results in \cref{fig:overall} and \cref{table:fulltable}, graph \GLsfull is a case that can show the overhead with a threshold $\beta = 8$.
as it has a large diameter and a small maximum $\maxcoreness$ value.

Although adversarial cases like above can be built,
the hybrid bucketing structure is still a good trade-off for most graphs in practice,
and it can be easily adjusted by changing the threshold $\beta$ for the triggering of the hierarchical bucketing structure.
}

\hide{
Our proposed algorithm framework is based on the peeling process,
which removes the vertices in the frontier in parallel.
At the same time, the degrees of the neighbors of the vertices in the frontier are decremented,
which may cause the neighbors to be added to the frontier for the next round or to be processed in the next few rounds.
Thus, it is worth considering the design of the data structure to maintain the vertices to save work for the synchronization in each round.

Bucketing is a common technique used in parallel algorithms to reduce the synchronization overhead of the parallel operations.
Because the neighbors of the vertices in the frontier will be decremented to a new degree value,
which may be the coreness value or close to the coreness value of the current round,
the vertices can be arranged into buckets based on their degree values.
In this way, we can save the synchronization work for check all the remaining vertices in each round.
For example, the figure in \cref{fig:eg-bucketing} shows the bucketing structure for the vertices in the frontier.
While processing the vertices in the frontier with coreness value $k = 3$,
multiple neighbors are moved between the buckets based on their decreased degree values,
or added to the buckets from the remaining bucket that contains the vertices with coreness value not in the current range of the buckets.
The buckets are processed with $k$ increasing, and will be synchronized once the buckets are all empty.

The implementation in \GBBS utilizes a multi-bucket data structure to map vertices to buckets based on their degree,
which is a straightforward way to reduce the synchronization overhead of the batch insertion operations (scan) for the buckets for each $k$-core round.
The advantage of the multi-bucket data structure is that it can reduce the synchronization overhead of the batch
insertion operations for the buckets for each $k$-core round.
Although the number of buckets can be determined by any fixed number,
it is a trade-off between the synchronization overhead and the space usage, as well as the frontier generating step mentioned in ~\ref{sec:framework}.
In comparison, previous algorithms take a single-bucket approach to store the vertices, i.e.,
there is only the frontier containing vertices that have the same coreness value $k$ in each round.
However, some graphs such as web networks and social networks have a very large maximum coreness value,
which may lead to a dominant synchronization overhead.

\subsection{Hierarchical Bucketing Structure}

To reduce the hashbag operations while arranging vertices into the buckets,
we design a hierarchical bucketing structure to balance the performance of the hashbag operations
and the space usage for the vertices movement while maintaining the work-efficiency of the algorithm.
For a graph with $n$ vertices,
we need to assign a space of $O(n)$ for each hashbag to store the vertices.
As we described in the previous subsection,
for dense graphs with larger numbers of peeling rounds,
opening more hashbags will increase the overhead of the hashbag operations,
as well as the space usage.
In contrast, for large-diameter graphs with fewer peeling rounds,
opening fewer hashbags will reduce the overhead of the hashbag operations.
To solve the unpredictable performance influenced by the factor of the number of peeling rounds,
we propose a hierarchical bucketing structure to organize the vertices.

Instead of assigning each bucket a single coreness value,
we assign a range of coreness values to each bucket,
which is increased exponentially.
Therefore, the number of buckets to maintain all the vertices is $\log n$,
which is space-efficient,
and the overhead of hashbag operations is reduced without setting a fixed number of buckets in advance.
The vertices are assigned to the buckets of the corresponding range based on their coreness values,
which can be calculated using bitwise operations.

\subsection{Bucketing Strategy}

}

\hide{

For instance, at the beginning of the algorithm, vertices are categorized into degrees of 0, 1, 2--3, 4--7, 8--15, ... .
In the first two rounds, we have $\frontier_0$ and $\frontier_1$ from the first two sets.
After that, we regenerate $\frontier_2$ and $\frontier_3$ from the third set with degrees 2--3.
On the fourth peeling round, vertices with degree 4--7 will be split into the first three sets, and the same process repeats.
We illustrate this process in \cref{fig:hier_bucketing}.

and the $i$-th set maintains vertices

The \HBS does not maintain the vertices with the same coreness value in a single bucket. Instead, it maintains the vertices with a range of coreness values in a single bucket.
Figure \ref{fig:hier_bucketing} shows an example of the hierarchical bucketing structure concept.
The coreness range of the vertices in each bucket is increased exponentially,
i.e., the first two buckets store the vertices with single coreness values (frontier buckets),
and from the third bucket, the coreness range is increased by  a factor of 2.
In the example of Figure \ref{fig:hier_bucketing}, the first bucket stores the vertices with coreness value 0, the second bucket stores the vertices with coreness value 1,
and the third bucket stores the vertices with coreness value 2 and 3, the range of the coreness value is increased by a factor of 2 for the next buckets.
\todo{}
In practice, we set an upper bound $F$ for the coreness range of all the buckets,
and the vertices with coreness value larger than $F$ will be maintained in a container for remaining vertices.

    \item \textbf{MakeBuckets($\mathbb{I}: identifiers$, \textsf{D[$\cdot$]} $\mapsto$ \textsf{bucket\_id})}: For an identifiers $i$ (vertex) in the remaining identifiers $\mathbb{I}$
        maintained outside of the buckets, map the identifiers (vertices) $i$ to the buckets with their corresponding range based on the $bucket\_id$ $D[i]$.
    \item \textbf{Insert($\textsf{D[i]} \mapsto \textsf{bucket\_id}$)}\\
        Insert a vertex to the bucket with the corresponding bucket id based on $d[i]$.

} 
\section{Experiments}\label{sec:exp}

\begin{table*}[!h] 
      \centering
      \small
      \vspace{-1em}
    \begin{tabular}{rccccc|rrr|r@{ }rrr|r}
          & \multicolumn{5}{c|}{\textit{\textbf{Graph Statistics}}} & \multicolumn{3}{c|}{\textit{\textbf{Ours}}} & \multicolumn{4}{c|}{\textit{\textbf{Baselines}}} & \multicolumn{1}{c}{\multirow{2}[1]{*}{\textit{\textbf{Notes}}}} \\
          & Name  & $n$   & $m$   & $\maxcoreness$ & $\rho$ & \multicolumn{1}{c}{seq.*} & \multicolumn{1}{c}{par.} & \multicolumn{1}{c|}{spd.} & \multicolumn{1}{c@{ }}{\textsf{BZ}*} & \multicolumn{1}{@{ }c}{\GBBS} & \multicolumn{1}{c}{\ParK} & \multicolumn{1}{c|}{\PKC} &  \\
\cmidrule{2-14}    \multicolumn{1}{c}{\multirow{5}[2]{*}{\begin{sideways}Social\end{sideways}}} & LJ    & \multicolumn{1}{r}{4.85M} & \multicolumn{1}{r}{85.7M} & \multicolumn{1}{r}{372} & \multicolumn{1}{r|}{3,480} & 2.37  & \textbf{.203} & 11.7  & 1.49  & .631  & .637  & .518  & \multicolumn{1}{l}{soc-LiveJournal1~\cite{backstrom2006group}} \\
          & OK    & \multicolumn{1}{r}{3.07M} & \multicolumn{1}{r}{234M} & \multicolumn{1}{r}{253} & \multicolumn{1}{r|}{5,667} & 3.94  & \textbf{.526} & 7.49  & 3.65  & 1.23  & 1.38  & .810  & \multicolumn{1}{l}{com-orkut~\cite{yang2015defining}} \\
          & WB    & \multicolumn{1}{r}{58.7M} & \multicolumn{1}{r}{523M} & \multicolumn{1}{r}{193} & \multicolumn{1}{r|}{2,910} & 29.5  & \textbf{.935} & 31.6  & 14.3  & 1.16  & 2.64  & 2.18  & \multicolumn{1}{l}{soc-sinaweibo~\cite{nr}} \\
          & TW    & \multicolumn{1}{r}{41.7M} & \multicolumn{1}{r}{2.41B} & \multicolumn{1}{r}{2,488} & \multicolumn{1}{r|}{14,964} & 62.2  & \textbf{2.72} & 22.9  & 61.2  & 4.79  & 857   & 75.6  & \multicolumn{1}{l}{Twitter~\cite{kwak2010twitter}} \\
          & FS    & \multicolumn{1}{r}{65.6M} & \multicolumn{1}{r}{3.61B} & \multicolumn{1}{r}{304} & \multicolumn{1}{r|}{10,034} & 126   & \textbf{3.68} & 34.3  & 174   & 6.18  & 416   & 33.1  & \multicolumn{1}{l}{Friendster~\cite{yang2015defining}} \\
\cmidrule{2-14}          & \multicolumn{5}{c|}{\textit{Geomean for Social Networks}} & \textit{18.5} & \textit{\textbf{.999}} &       & \textit{15.3} & \textit{1.93} & \textit{15.3} & \textit{4.70} &  \\
\cmidrule{2-14}    \multicolumn{1}{c}{\multirow{5}[2]{*}{\begin{sideways}Web\end{sideways}}} & EH    & \multicolumn{1}{r}{11.3M} & \multicolumn{1}{r}{522M} & \multicolumn{1}{r}{9,877} & \multicolumn{1}{r|}{7,393} & 8.21  & \textbf{.795} & 10.3  & 5.49  & 1.39  & 5.67  & 8.22  & \multicolumn{1}{l}{eu-host~\cite{webgraph}} \\
          & SD    & \multicolumn{1}{r}{89.3M} & \multicolumn{1}{r}{3.88B} & \multicolumn{1}{r}{10,507} & \multicolumn{1}{r|}{19,063} & 140   & \textbf{4.39} & 32.0  & 143   & 6.56  & 410   & 57.5  & \multicolumn{1}{l}{sd-arc~\cite{webgraph}} \\
          & CW    & \multicolumn{1}{r}{978M} & \multicolumn{1}{r}{74.7B} & \multicolumn{1}{r}{4,244} & \multicolumn{1}{r|}{106,819} & 2453  & \textbf{28.6} & 85.8  & 2328 & 53.2  & \textit{T/O} & \textit{T/O} & \multicolumn{1}{l}{ClueWeb~\cite{webgraph}} \\
          & HL14  & \multicolumn{1}{r}{1.72B} & \multicolumn{1}{r}{124B} & \multicolumn{1}{r}{4,160} & \multicolumn{1}{r|}{58,737} & 3587  & \textbf{54.7} & 65.5  & \textit{OOM} & 72.0  & \textit{OOM} & \textit{OOM} & \multicolumn{1}{l}{Hyperlink14~\cite{webgraph}} \\
          & HL12  & \multicolumn{1}{r}{3.56B} & \multicolumn{1}{r}{226B} & \multicolumn{1}{r}{10,565} & \multicolumn{1}{r|}{130,737} & 9177  & \textbf{108} & 84.6  & \textit{OOM} & 152   & \textit{OOM} & \textit{OOM} & \multicolumn{1}{l}{Hyperlink12~\cite{webgraph}} \\
\cmidrule{2-14}          & \multicolumn{5}{c|}{\textit{Geomean for Web Networks}} & \textit{622} & \textit{\textbf{14.3}} &       & \textit{N/A} & \textit{22.1} & \textit{N/A} & \textit{N/A} &  \\
\cmidrule{2-14}    \multicolumn{1}{c}{\multirow{4}[2]{*}{\begin{sideways}Road\end{sideways}}} & AF    & \multicolumn{1}{r}{33.5M} & \multicolumn{1}{r}{88.9M} & \multicolumn{1}{r}{3} & \multicolumn{1}{r|}{189} & 9.83  & \textbf{.155} & 63.2  & 5.54  & .281  & .363  & .253  & \multicolumn{1}{l}{OSM Africa~\cite{roadgraph}} \\
          & NA    & \multicolumn{1}{r}{87.0M} & \multicolumn{1}{r}{220M} & \multicolumn{1}{r}{4} & \multicolumn{1}{r|}{286} & 32.4  & .432  & 74.8  & 12.4  & .682  & .724  & \textbf{.417} & \multicolumn{1}{l}{OSM North America~\cite{roadgraph}} \\
          & AS    & \multicolumn{1}{r}{95.7M} & \multicolumn{1}{r}{244M} & \multicolumn{1}{r}{4} & \multicolumn{1}{r|}{343} & 34.8  & \textbf{.480} & 72.5  & 16.0  & .709  & .878  & .656  & \multicolumn{1}{l}{OSM Asia~\cite{roadgraph}} \\
          & EU    & \multicolumn{1}{r}{131M} & \multicolumn{1}{r}{333M} & \multicolumn{1}{r}{4} & \multicolumn{1}{r|}{513} & 47.4  & .679  & 69.8  & 33.2  & .925  & .869  & \textbf{.609} & \multicolumn{1}{l}{OSM Europe~\cite{roadgraph}} \\
\cmidrule{2-14}          & \multicolumn{5}{c|}{\textit{Geomean for Road Networks}} & \textit{26.9} & \textit{\textbf{.385}} &       & \textit{13.8} & \textit{.595} & \textit{.669} & \textit{.453} &  \\
\cmidrule{2-14}    \multicolumn{1}{c}{\multirow{5}[2]{*}{\begin{sideways}\KNN\end{sideways}}} & CH5   & \multicolumn{1}{r}{4.21M} & \multicolumn{1}{r}{29.7M} & \multicolumn{1}{r}{5} & \multicolumn{1}{r|}{7} & .826  & \textbf{.021} & 39.1  & .431  & .042  & .037  & \textbf{.021} & \multicolumn{1}{l}{Chem~\cite{fonollosa2015reservoir,wang2021geograph}, $k=5$} \\
          & GL2   & \multicolumn{1}{r}{24.9M} & \multicolumn{1}{r}{65.3M} & \multicolumn{1}{r}{2} & \multicolumn{1}{r|}{12} & 6.96  & \textbf{.109} & 64.1  & 7.69  & .167  & .155  & .113  & \multicolumn{1}{l}{GeoLife~\cite{geolife,wang2021geograph}, $k=2$} \\
          & GL5   & \multicolumn{1}{r}{24.9M} & \multicolumn{1}{r}{157M} & \multicolumn{1}{r}{5} & \multicolumn{1}{r|}{42} & 6.81  & \textbf{.125} & 54.7  & 3.54  & .196  & .179  & .249  & \multicolumn{1}{l}{GeoLife~\cite{geolife,wang2021geograph}, $k=5$} \\
          & GL10  & \multicolumn{1}{r}{24.9M} & \multicolumn{1}{r}{310M} & \multicolumn{1}{r}{10} & \multicolumn{1}{r|}{16} & 8.46  & \textbf{.162} & 52.4  & 5.57  & .277  & .175  & .168  & \multicolumn{1}{l}{GeoLife~\cite{geolife,wang2021geograph}, $k=10$} \\
          & COS5  & \multicolumn{1}{r}{321M} & \multicolumn{1}{r}{1.96B} & \multicolumn{1}{r}{2} & \multicolumn{1}{r|}{23} & 117   & \textbf{2.06} & 56.6  & 61.9  & 3.66  & 2.74  & 2.08  & \multicolumn{1}{l}{Cosmo50~\cite{cosmo50,wang2021geograph}, $k=5$} \\
\cmidrule{2-14}          & \multicolumn{5}{c|}{\textit{Geomean for \KNN Graphs}} & \textit{8.27} & \textit{\textbf{.157}} &       & \textit{5.26} & \textit{.268} & \textit{.218} & \textit{.183} &  \\
\cmidrule{2-14}    \multicolumn{1}{c}{\multirow{6}[2]{*}{\begin{sideways}Others\end{sideways}}} & TRCE & \multicolumn{1}{r}{16.0M} & \multicolumn{1}{r}{48.0M} & \multicolumn{1}{r}{2} & \multicolumn{1}{r|}{1,839} & 2.03  & \textbf{.066} & 31.0  & 1.49  & 1.96  & .424  & .067  & \multicolumn{1}{l}{Huge traces~\cite{nr}} \\
          & BBL & \multicolumn{1}{r}{21.2M} & \multicolumn{1}{r}{63.6M} & \multicolumn{1}{r}{2} & \multicolumn{1}{r|}{1,915} & 3.18  & \textbf{.077} & 41.1  & 3.36  & 1.80  & .203  & .081  & \multicolumn{1}{l}{Huge bubbles~\cite{nr}} \\
          & GRID  & \multicolumn{1}{r}{100M} & \multicolumn{1}{r}{400M} & \multicolumn{1}{r}{2} & \multicolumn{1}{r|}{50,499} & 6.21  & \textbf{.282} & 22.1  & 14.1  & 14.8  & 8.03  & 3.21  & \multicolumn{1}{l}{Synthetic grid graph} \\
          & CUBE  & \multicolumn{1}{r}{1.00B} & \multicolumn{1}{r}{6.0B} & \multicolumn{1}{r}{3} & \multicolumn{1}{r|}{2,895} & 183   & \textbf{4.01} & 45.7  & 162   & 9.46  & 110   & 10.8  & \multicolumn{1}{l}{Synthetic cubic graph} \\
          & HCNS  & \multicolumn{1}{r}{0.1M} & \multicolumn{1}{r}{5.0B} & \multicolumn{1}{r}{50,000} & \multicolumn{1}{r|}{50,000} & 27.8  & \textbf{2.01} & 13.8  & 23.5  & 16.0  & 49.7  & \textit{OOM} & \multicolumn{1}{l}{Synthetic high-coreness graph} \\
          & HPL   & \multicolumn{1}{r}{100M} & \multicolumn{1}{r}{1.20B} & \multicolumn{1}{r}{3,980} & \multicolumn{1}{r|}{6,297} & 47.3  & \textbf{1.77} & 26.8  & 38.9  & 3.59  & 30.4  & 59.1  & \multicolumn{1}{l}{Synthetic power-law graph} \\
\cmidrule{2-14}          & \multicolumn{5}{c|}{\textit{Geomean for Other Graphs}} & \textit{14.6} & \textit{\textbf{.523}} &       & \textit{14.8} & \textit{5.52} & \textit{6.97} & \textit{N/A} &  \\
\cmidrule{2-14}    \end{tabular}%

  \caption{\textbf{Graph information and running time (in seconds) of tested algorithms.} 
  \textnormal{
    $n= |V|$, $m= |E|$.
    $\maxcoreness=$ maximum coreness value. 
    $\rho$ = the peeling complexity~\cite{dhulipala2017}, i.e., the number of subrounds using offline peeling. 
    ``seq.'' $=$ our sequential running time, ``par.'' $=$ our parallel running time.
    ``spd.'' $= \text{seq.} / \text{par.}$, i.e., self-relative speedup of our algorithm. 
    Baselines include the \BZ{} algorithm~\cite{batagelj2003m}, \Julienne{}~\cite{dhulipala2017,gbbs2021},
    \Park{}~\cite{dasari2014park} and \PKC~\cite{kabir2017parallel}. 
    ``*'' indicates the algorithm is sequential.
    The best running time on each graph is highlighted in bold.
    ``T/O'': timeout ($>2000$ seconds) for parallel algorithms.
    ``OOM'': out of memory.
  }
      \label{table:fulltable}
      }
    
    \end{table*}%

\subsection{Experiment Setup}
Our experiments were conducted on a 96-core machine (192 hyper-threads) equipped with four 2.1 GHz Intel Xeon Gold 6252 CPUs, 
each with a 36MB L3 cache and 1.5TB of main memory. 
For all tests, except the sequential ones, 
we used \texttt{numactl -i all} to interleave memory across all CPUs. 
We report the average runtime of five runs, following an initial warm-up run.

\subsubsection{Datasets}\label{sec:datasets}
We tested a wide range of real-world graphs, including large-scale social networks, web graphs, 
road networks, \knn{} graphs, and various other graphs.
Additionally, we generate synthetic graphs to simulate adversarial scenarios for both existing baselines and our own algorithm. 
Directed graphs are symmetrized by converting edges to bidirectional.
The graph information, along with their references and acronyms, are shown in \cref{table:fulltable}, 
and we will refer to the graphs using these acronyms throughout the rest of this section.
\revise{
    The $k$-NN graphs are generated from a set of vectors (multi-dimensional points). 
    Each vertex represent a point (vector), and it has directed edges to its $k$ nearest neighbors. 
    The five graphs in our experiments are from real-world vector datasets~\cite{fonollosa2015reservoir,wang2021geograph,geolife,cosmo50}.
}
\TRCE and \BBL are meshes taken from individual frames of sequences that resembles two-dimensional adaptive numerical simulations~\cite{nr}.
The two graphs \GRD{} and \CBC{} are a $10^4 \times 10^4$ 2D grid and a $10^3 \times 10^3 \times 10^3$ 3D cube, respectively.
\HCNS{} is a synthetic graph with a high $\maxcoreness=50000$. 
It contains exactly one vertex with coreness $i$ for $1\le i<\kmax$, and a dense subgraph with coreness $\maxcoreness$.  
\HPL{} is a power-law degree distribution graph, generated using the Barabasi-Albert model~\cite{barabasi1999emergence}.
We classify the social networks, web graphs, \HCNS{}, and \HPL{} as \emph{dense graphs} due to their relatively large average degrees and high coreness, 
while the remaining graphs are categorized as \emph{sparse graphs}.

\subsubsection{Baselines}
We compare our algorithm against three state-of-the-art parallel baselines: \Julienne~\cite{dhulipala2017}, \ParK~\cite{cheng2013local}, \PKC~\cite{kabir2017parallel}.
Each algorithm exhibits specific strengths and weaknesses depending on the graph types. 
These algorithms are discussed in more detail in \cref{sec:peeling}. 
We also compare our algorithm with the sequential algorithm \BZ{}. 
The implementation for \Park{}, \PKC{}, and \BZ{} are from the code of the \PKC{} paper~\cite{kabir2017parallel}.
The implementation for \Julienne{} is from the GBBS library~\cite{gbbs2021}, in which \Julienne{} is integrated. 


\subsection{Experimental Results}
\begin{figure*}[t]
  \centering
  \includegraphics[width=\textwidth]{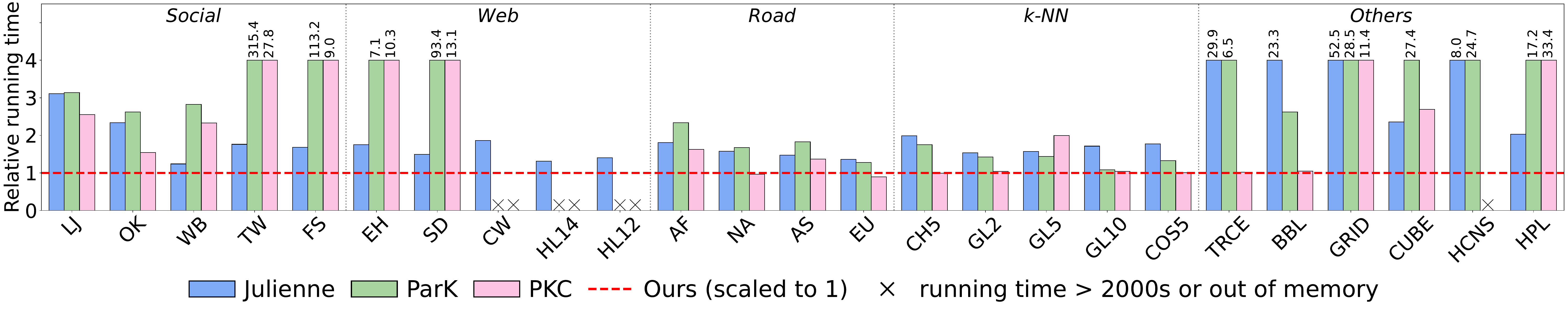}
  \caption{\textbf{Relative running time of \ParK~\cite{dasari2014park}, \PKC~\cite{kabir2017parallel} and \GBBS~\cite{dhulipala2017, gbbs2021} normalized to our running time (red dotted line) on all graphs. Lower is better.}
  The bars are truncated at 4 for better visualization. The text on the bars are actual relative running time.
  }\label{fig:overall}
\end{figure*} 

  \begin{figure*}[htpb]
  \centering
  \includegraphics[width=\textwidth]{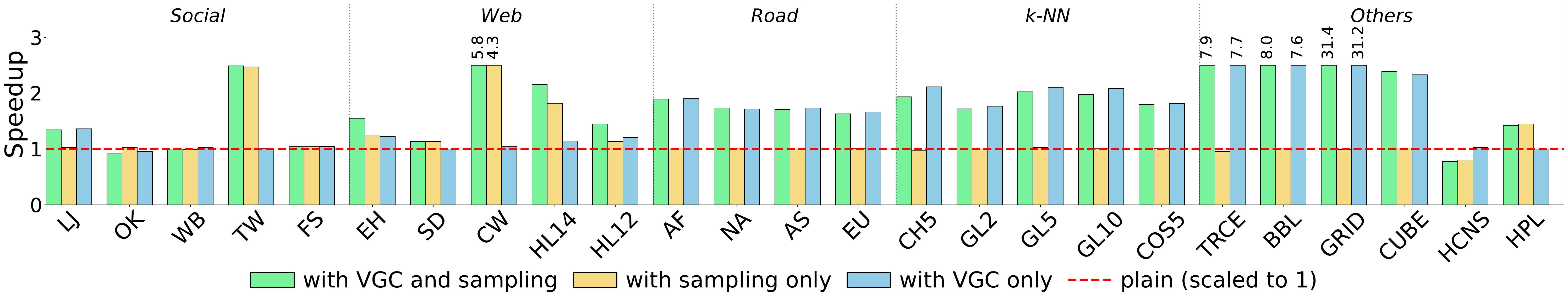}
  \caption{\textbf{Speedup of VGC and sampling over a plain implementation. Higher is better.}
  The plain version does not use VGC or sampling. 
  The bars are truncated at 2.5 for better visualization. The text on the bars are the actual speedup values.
  }\label{fig:self_comparison}
\end{figure*}

\subsubsection{Overall Performance}
We present the overall performance of our algorithm and all baselines in \cref{table:fulltable}, 
and the relative running times of all parallel baselines (normalized to ours) in \cref{fig:overall}. 
In \cref{table:fulltable}, 
we highlight the best running time for each graph in bold. 
We also report the sequential running time of our algorithm along with the self-relative speedup. 
In most cases, our sequential runtime is comparable to or better than \BZ{}, 
indicating the work-efficiency of our approach.
Our algorithm also demonstrates significant parallelism, 
achieving a self-relative speedup of 7.5--86$\times$. 
In contrast,
all baselines exhibit unsatisfactory performance on some graphs due to the lack of parallelism.
On certain graphs, they perform even slower than the sequential implementations (\BZ{} or our sequential time) and much slower than other baselines. 
Our algorithm consistently outperforms the best sequential time by 6.9--85$\times$. 
Combining both work-efficiency and strong parallelism, 
our algorithm is the fastest on 23 out of 25 tested graphs, except for \EU{} and \NA{}, 
where it remains competitive and only <12\% slower than the best baseline.

\cref{fig:overall} presents relative running time of all parallel implementations, normalized to ours. 
Each baseline may exhibit tens of times slowdown than our algorithm on certain graphs. 
Such performance degeneration occurs on different sets of graphs for each baseline, due to the different design of each algorithm. 

All algorithms perform well on \knn{} graphs, due to their properties: all vertices have small degrees, 
the same coreness, and only require a few subrounds to complete.
Despite small variation in performance, our algorithm is slightly faster on average.

For dense graphs (social networks, web graphs, \HCNS{} and \HPL{}),
which contain many high-degree vertices,
\Julienne{} performs well due to work-efficiency and their race-free offline peeling algorithm. 
\PKC{} and \Park{} have much worse performance
due to 1) work-inefficiency from not maintaining the active set and 2) high contention when updating the \induceddegree{s} of the vertices.  
Our algorithm outperforms \Julienne{}'s offline algorithm by 1.24--3.11$\times$, due to work-efficiency and the sampling scheme that reduces contention during the online peeling process. 

On sparse graphs, however, \Julienne{} may have poor performance due to the overhead of enabling race-freedom and fully synchronized subrounds in the offline algorithm. 
In this case, the simpler online algorithm in \PKC{} and \Park{} performs better since very light contention is incurred. 
\PKC{}'s thread-local buffer also fully avoids subrounds, and thus it achieves the best performance on three sparse graphs. 
However, despite being the fastest on three sparse graphs, \PKC{} also leads to the most timeout or out-of-memory cases in our test, and may have much worse performance than others on dense graphs.  
This highlights the intrinsic difficulty to optimize the performance of parallel \kcore{} algorithms: 
employing an optimization may address issues on specific graphs, but may sacrifice the performance on other graphs. 

The consistent good performance of our algorithm across diverse graph types is enabled by our new techniques. 
Sampling optimizes for high-degree vertices in dense graphs, 
while VGC is particularly effective for low-degree vertices in sparse graphs.
Consequently, our algorithm consistently performs well on various graph types, 
and prevents significant performance drops on any type of graphs.

\subsubsection{Evaluation on VGC and Sampling}

\afterpage{
  \begin{figure}[htbp]
  \centering
  \includegraphics[width=\columnwidth]{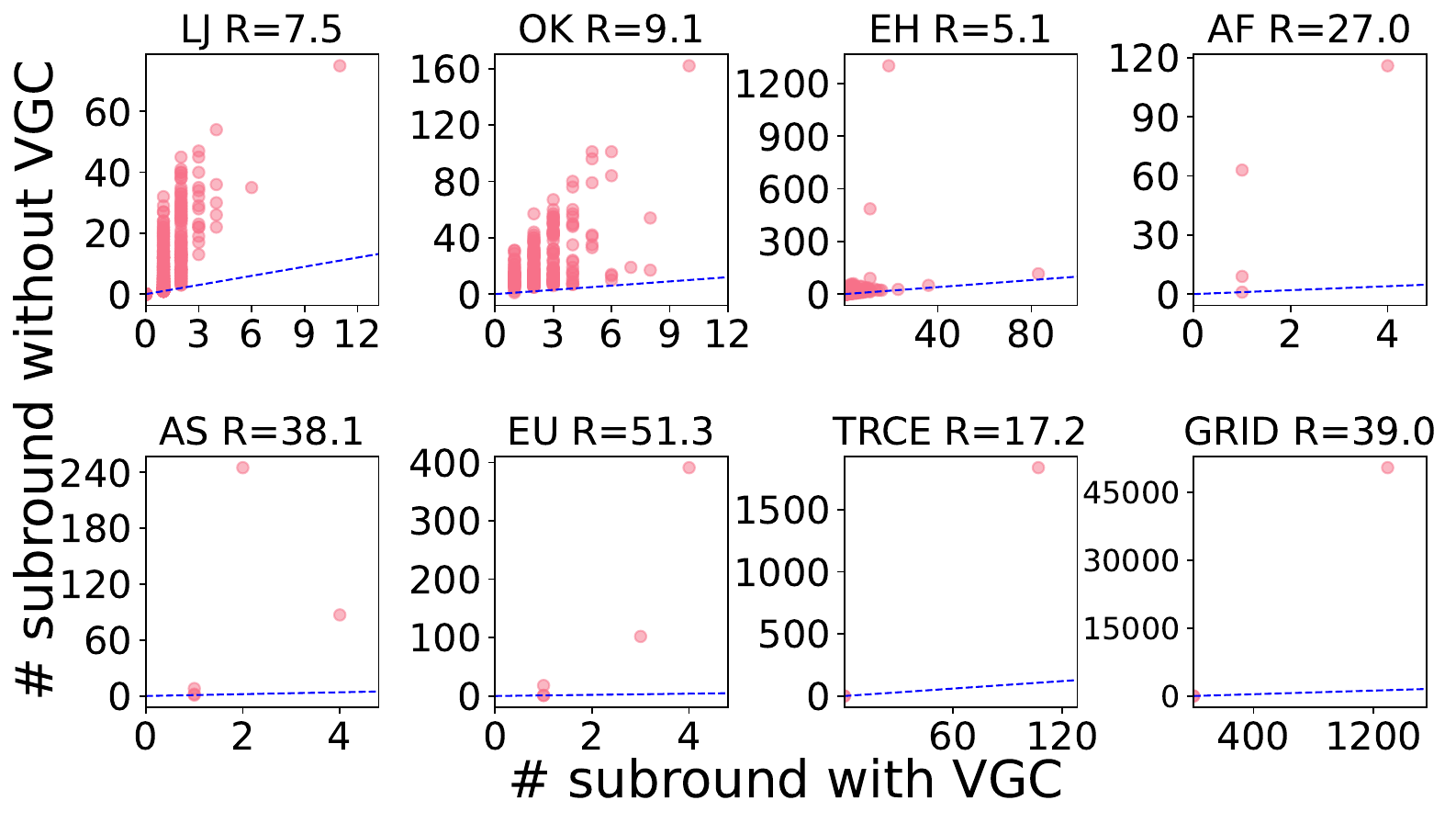}
  \caption{\textbf{Number of subrounds with and without VGC.} 
  Each point $(x,y)$ means VGC reduces a round with $y$ subrounds to $x$ subrounds. 
  The blue dotted line is the baseline where $y=x$. 
  The number $R$ in each subtitle is the reduction ratio of the number of subrounds.
  }\label{fig:queue-rho}
\end{figure}
}

In this section, we evaluate the effectiveness of our online framework using VGC and sampling schemes. 
We compare four versions of our algorithm:
one without sampling or VGC (referred to as the \emph{plain} version), 
one using only sampling, 
one using only VGC, 
and our final with both sampling and VGC.
We show the speedup of the latter three over the plain version in \cref{fig:self_comparison}. 
As discussed in \cref{sec:alg}, 
these two techniques are optimized for different scenarios: 
sampling enhances performance for high-degree vertices, 
while VGC is designed to optimize for low-degree vertices. 
In practice, most graphs benefit primarily from one of these techniques. 
Nearly all sparse graphs show significant improvement with VGC.
Seven dense graphs benefit from sampling.
On all graphs but \HCNS{}, using sampling and/or VGC improves the performance over the plain version. 

We first study the impact of sampling. 
In fact, eight graphs (\TW{}, \EH{}, \SD{}, \CW{}, \HL{}, \HLs{}, \HPL{}, and \HCNS{}) contain vertices with very high degrees and trigger sampling. 
\revise{
    \iffullversion{We show the sampling speedup for these eight graphs in \cref{fig:exp-sampling}.}
}
Seven of them (all except \HCNS{}) benefit from it. 
We first discuss the adversarial graph \HCNS{} with $\kmax=50000$. 
In this case, sampling slightly decreases the performance by 24\%. 
The overhead comes from the validation step in each round which processes all vertices in sample mode. 
On \HCNS{}, half of the vertices (all those in $\kmax$-core) need to be checked until $k=\Theta(\kmax)$.  
On real-world graphs that trigger sampling (e.g., the power-law graphs), 
this overhead is minimal since only a small number of vertices have high degrees and enter sampling mode.
Therefore, this adversarial graph \HCNS{} roughly illustrate the cost of sampling, which is reasonably low (about 24\%). 
For the other seven graphs that trigger sampling, they all benefit from it by up to 4.3$\times$ (on \CW{}).
In fact, while only a small fraction of vertices may require sampling, the benefit is significant. 
For example, on \TW{}, although only about 1000 out of 40 million vertices are in sample mode, the improvement is 2.4$\times$. 
This indicates that a small fraction of high-degree vertices lead to high contention that bottlenecks the entire computation.
Our sampling scheme effectively mitigates this contention, resulting in much improved performance.

For VGC, all sparse graphs and certain dense graphs benefit from it, with a speedup of up to 31.2$\times$ (on \GRD{}). 
We did not observe notable performance drop due to VGC, which indicates that the overhead of VGC is negligible. 
To further study the impact of VGC, we compare the number of subrounds with and without VGC on some representative graphs
in \cref{fig:queue-rho}. 
For each red point in \cref{fig:queue-rho}, the y-coordinate is the number of subrounds without VGC, 
and the x-coordinate is the number of subrounds using VGC. 
We can see that the number of subrounds is significantly reduced on various graphs.
Even on dense graphs, the numbers of subrounds also decrease by up to 9.1$\times$ (on \OK{}). 
However, the improvement in time is not as large as sparse graphs, 
because the computation on dense graphs is also substantial, and the scheduling overhead does not dominate the time.
Therefore, optimizing this part by VGC only marginally improves the performance for dense graphs. 
On sparse graphs, VGC is very effective. 
On road networks, the numbers of subrounds in each round are reduced from hundreds to within 4, with an improvement of 26--51$\times$. 
This also leads to 1.7--1.9$\times$ improvement in time. 
The improvement is more significant on four other graphs \TRCE{}, \BBL{}, \GRD{}, and \CBC{}, 
which simulates extreme cases for sparse graphs. 
The reduce ratio of \subround{s} ranges from 10--39$\times$, 
resulting in a speedup of 2.3--31.2$\times$ compared to the plain algorithm.

\subsubsection{Evaluation of \HBS} \label{para:exp_bucketing}

\begin{figure}[t]
  \centering
  \includegraphics[width=\columnwidth]{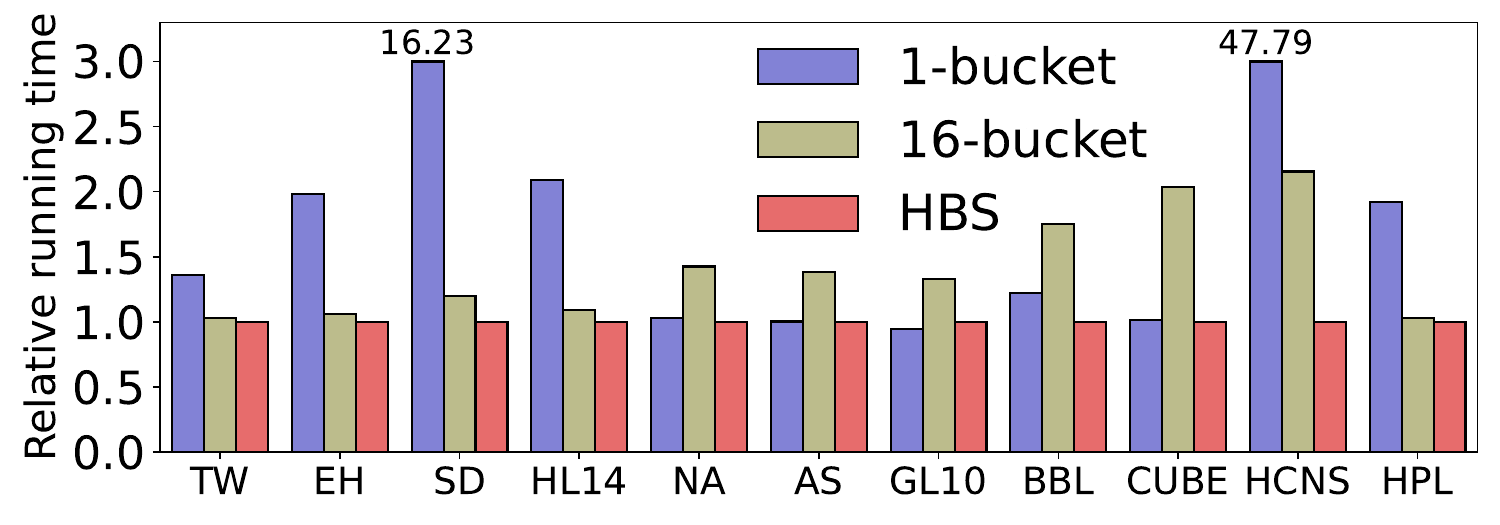}
  \caption{\textbf{
  Relative running time of different bucketing strategies, normalized to \HBS{}. Lower is better.}
  }\label{fig:bucketing-exp}
\end{figure} 
\revise{\begin{figure*}[h!]
  \centering
  \includegraphics[width=\textwidth]{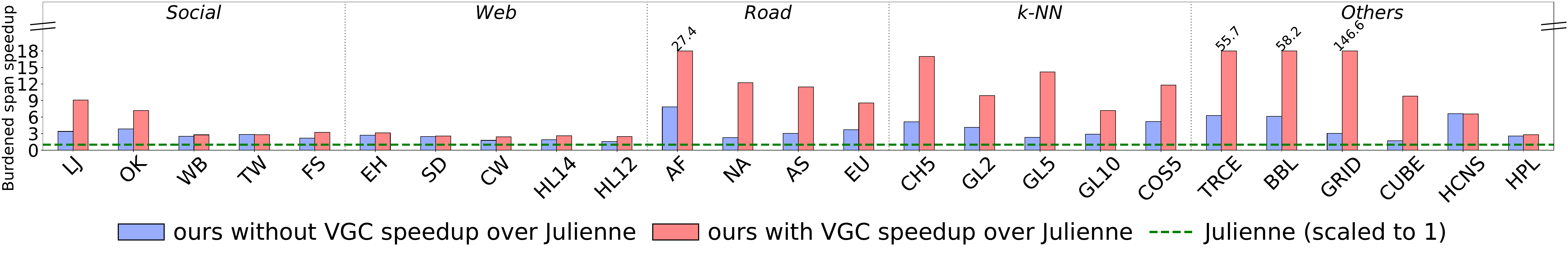}
  \caption{\textbf{Burdened span~\cite{he2010cilkview} speedup of our algorithm (with and without VGC) over \GBBS~\cite{dhulipala2017, gbbs2021} (red dotted line, always$=1$) on all graphs. Higher is better.}
  The bars are truncated at 18 for better visualization. The text on the bars are actual burdened span speedup.
  \label{fig:burdened_span}}
\end{figure*} }
We evaluate the effectiveness of \HBS{} by comparing it to two baselines: 
always using one bucket, as shown in the original framework (\cref{algo:framework}), 
and always using 16 buckets, similar to \Julienne{}'s strategy.
\revise{
    Different from the other two technique (VGC and sampling), \HBS{} can be applied to both the online and offline peeling algorithms, and therefore
    we analyze it separately here. 
    \ifconference{For completeness, we also analyze the performance of all possible combinations of the three techniques in the full version of this paper.}
    \iffullversion{For completeness, we also analyze the performance of all possible combinations of the three techniques in \cref{table:combinations}.}
}
\cref{fig:bucketing-exp} shows the relative running times (normalized to our method) across representative graphs.
Using one bucket involves constructing the bucket every round, 
and thus has low performance on dense graphs with large average degrees.
Using 16 buckets means to extract and partition vertices into buckets every 16 rounds, 
which is more efficient on dense graphs. 
However, as discussed in \cref{sec:bucketing}, on graphs with a constant average degree, 
using bucket structures does not theoretically improve the cost.
The overhead of managing the bucketing structure can make it much slower than using 1 bucket. 

Recall that our \HBS{} start with a single bucket for sparse graphs with an average degree of 16 or less,
and start to maintain a hierarchy of buckets when $k$ reaches 16, the same threshold.
This self-adaptive strategy adjusts to the density of the graph.
Across all graphs, \HBS{} matches the performance of the better option between one or 16 buckets, 
and in some cases, performs much faster than both.
For dense graphs, our hierarchical structure is always as good as 16-bucket, 
and has superior performance on very dense graphs. On the extreme case \HCNS{}, using a \HBS{} is
$47.8\times$ faster than 1-bucket and $2.01\times$ than 16-bucket structure.


\subsubsection{Scalability}
\begin{figure}[t]
  \centering
  \includegraphics[width=\columnwidth]{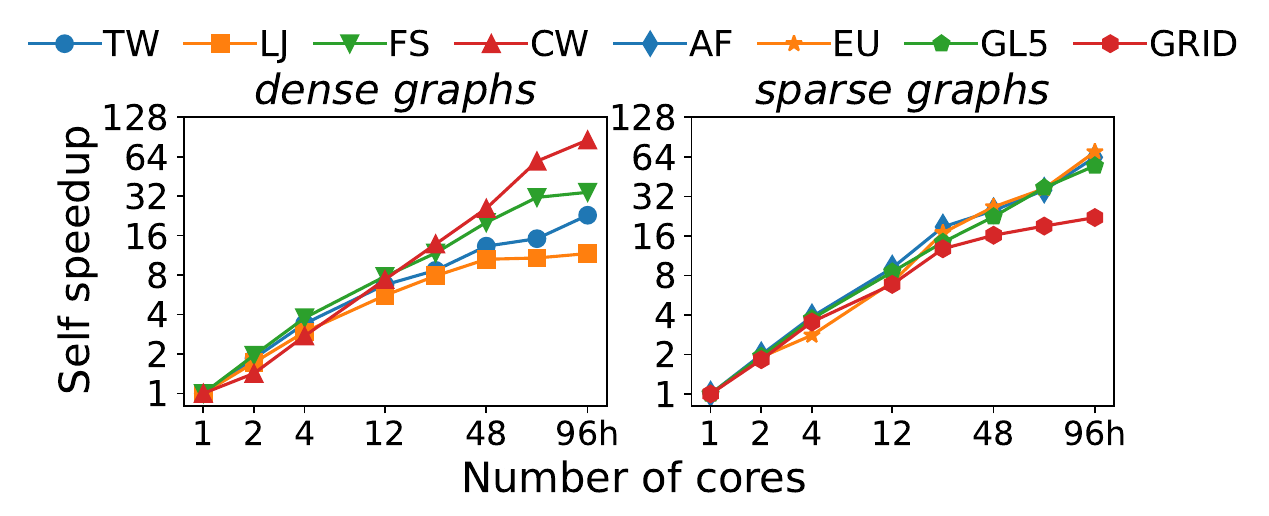}
  \caption{\textbf{Self-relative speedup on dense graphs and sparse graphs.}
  ``96h'' = 96 cores with hyperthreading.
  }\label{fig:scalability}
\end{figure}

The two techniques proposed in \cref{sec:alg} both aim to improve parallelism. 
To evaluate this, we test the scalability of our algorithm on dense and sparse graphs, respectively, in \cref{fig:scalability}. 
The self-relative speedup for all graphs is also reported in \cref{table:fulltable}. 
In general, the scalability is more significant on larger graphs, 
since there are potentially more computation to exploit parallelism. 
With our new techniques, most sparse graphs achieve a self-relative speedup of 50$\times$ or more. 
For dense graphs, the two graphs (\CW{} and \HL{}) achieves self-speedup exceeding 80$\times$. 
Interestingly, we observe that these two graphs can take advantage from both sampling and VGC (see \cref{fig:self_comparison}). 
This further indicates the effectiveness of our new techniques in improving parallelism. 

\revise{
\subsubsection{Burdened Span Analysis}
\label{sec:exp:span}
    To evaluate how VGC reduces the burdened span and thus improves the performance, 
    we use Cilkview~\cite{he2010cilkview} to measure the burdened span of different algorithms in practice. 
    Cilkview only applies to algorithms in OpenCilk~\cite{opencilk}, and 
    both our code and \Julienne{} can be easily compiled with OpenCilk. 
    We run our algorithm with and without VGC, as well as Julienne~\cite{dhulipala2017}, on all graphs.
\cref{fig:burdened_span} shows the speedup of our algorithm (with and without VGC) over \Julienne{} (higher is better), and the green dotted line at 1 represent the burdened span of \Julienne{}. 

As shown in \cref{fig:burdened_span},
the burdened span of our plain algorithm (without VGC) is a constant factor (1.6--7.9$\times$) better than that of Julienne~\cite{dhulipala2017},
primarily due to the simplicity of the online algorithm.
Since \Julienne{} is offline, running histogram and semisort incurs additional rounds of global synchronization. 
This gap, although as a constant factor in theory, slightly differs the measured burdened span and explains the performance differences. 

VGC further reduces the burdened span, particularly on sparse graphs, achieving up to 147$\times$ improvement over \Julienne{}.
Combining with the results in \cref{fig:overall}, the speedup of burdened span matches the speedup of the running time of the algorithms. 
For example, the three most significant burdened span improvements with VGC are on \TRCE{}, \BBL{}, and \GRD{}, which also show the best time speedups over \Julienne{}.
Such highly correlated trends can be observed on other graphs. 
This suggests that the reduction in burdened span (i.e., reducing the synchronization overhead) is the primary reason for our algorithm to outperform \Julienne{}. 
}

\subsubsection{Additional Experiments} 
In the full paper, 
we also evaluate our algorithm on the task of computing a specific \kcore{} of a graph given a particular value of $k$, and compare it with a graph library Galois~\cite{Nguyen2014}. 
With different values of $k$, on two graphs \OK{} and \TW{}, our algorithm outperforms Galois by 1.6--6.2$\times$.

\section{Related Work}
The $k$-core decomposition problem has been extensively studied since the introduction by Seidman~\cite{seidman1983network}.
The first sequential algorithm was proposed by Matula and Beck~\cite{matula1983smallest},
using a bucket sort to arrange vertices by degree and iteratively deleting vertices with degree $k$ (peeling) until all are removed.
The algorithm runs in $O(m + n)$.
Batagelj and Zaversnik (\BZ)~\cite{batagelj2003m} provided a sequential implementation with the same time complexity. 

The \kcore{} problem is also carefully studied in the shared-memory parallel setting~\cite{dasari2014park,kabir2017parallel,dhulipala2017,montresor2011distributed,khaouid2015k,li2021k}, as well as included in many parallel graph libraries such as GraphX~\cite{gonzalez2014graphx}, Powergraph~\cite{gonzalez2012powergraph}, Ligra~\cite{shun2013ligra} and Julienne~\cite{dhulipala2017} (later integrated into the GBBS library~\cite{gbbs2021}).
In this paper, we compared to the three state-of-the-part solutions, including \ParK~\cite{dasari2014park}, \PKC~\cite{kabir2017parallel}, and \Julienne{}~\cite{dhulipala2017}.
Among them, \ParK and \PKC use the online peeling process, and \Julienne is offline.
More details about them were overviewed in \cref{sec:peeling}.

Given the wide applicability, \kcore is also extensively studied in other settings, such as on GPUs~\cite{mehrafsa2020vectorising, ahmad2023accelerating, zhao2024pico, li2021k, tripathy2018scalable, zhao2024speedcore, zhang2017accelerating, wang2016gunrock}, the external memory (disk) setting~\cite{cheng2011efficient, wen2018efficient}, and the low-memory setting~\cite{khaouid2015k}.
The techniques proposed in these papers mostly focused on the specific challenges in each setting, and have small overlaps with the new techniques introduced in this paper.

We are also aware of variants of $k$-core decomposition.
A direct extension is to maintain \kcore with vertex and edge updates.
Research has been done on this topic, for both the dynamic (online) setting~\cite{aksu2014distributed, aridhi2016distributed, gabert2022batch,liu2022parallel} and the streaming (offline) setting~\cite{sariyuce2013streaming, esfandiari2018parallel, sariyuce2016incremental}.
Another direction is computing approximate $k$-core decomposition, both in sequential settings~\cite{king2022computing} 
and parallel settings~\cite{esfandiari2018parallel,liu2022parallel, liu2024parallel, dhulipala2022differential}.
On directed graphs, s similar problem is referred to as $D$-core decomposition.
Algorithms for it have also been studied recently~\cite{luo2024efficient, liao2022distributed, giatsidis2013d}.
The $k$-core decomposition also relates to many other problems, 
such as dense subgraph discovery~\cite{luo2023scalable} and hierarchical core decomposition~\cite{chu2022hierarchical}.
How to apply our new techniques in these related problems can be an interesting future work.

\section{Conclusion} 
In this paper, we present an efficient parallel solution to \kcore{} decomposition.
Our contributions include 1) an algorithmic framework that can achieve work-efficiency using various peeling strategies,
2) a sampling scheme to reduce contention on high-degree vertices on dense graphs,
3) a vertical granularity control (VGC) algorithm to hide the scheduling overhead for low-degree vertices on sparse graphs,
and 4) a hierarchical bucketing structure (HBS) to improve the performance of managing frontiers. 
With the new techniques, our algorithm achieves high performance on a variety of graphs. 
We compare our algorithm with three state-of-the-art algorithm \Julienne{}, \Park{}, and \PKC{}.
Due to different design, each baseline may exhibit worse performance than a sequential implementation on certain graphs. 
Our algorithm consistently performs well on all graphs, and always outperforms the best sequential algorithm by 7.3-84 times. 
In the experiment, our algorithm is faster than the baselines on all 12 dense graphs and 11 out of 13 sparse graphs. 
Experimental results also show that each of our new technique is effective in improving performance. 
\section{Acknowledgement}\label{sec:ack}
This work is supported by NSF grants CCF-2103483, TI-2346223, IIS-2227669, NSF CAREER Awards CCF-2238358 and CCF2339310, 
the UCR Regents Faculty Development Award, and the
Google Research Scholar Program.
\bibliographystyle{ACM-Reference-Format}
\balance
\bibliography{bib/strings, bib/main}

\iffullversion{
  \clearpage
  \appendix
\clearpage
\label{sec:appendix}


\revise{
\section*{A.~~Speedup with and without Sampling}

\begin{figure}[!t]
  \centering
  \includegraphics[width=\columnwidth]{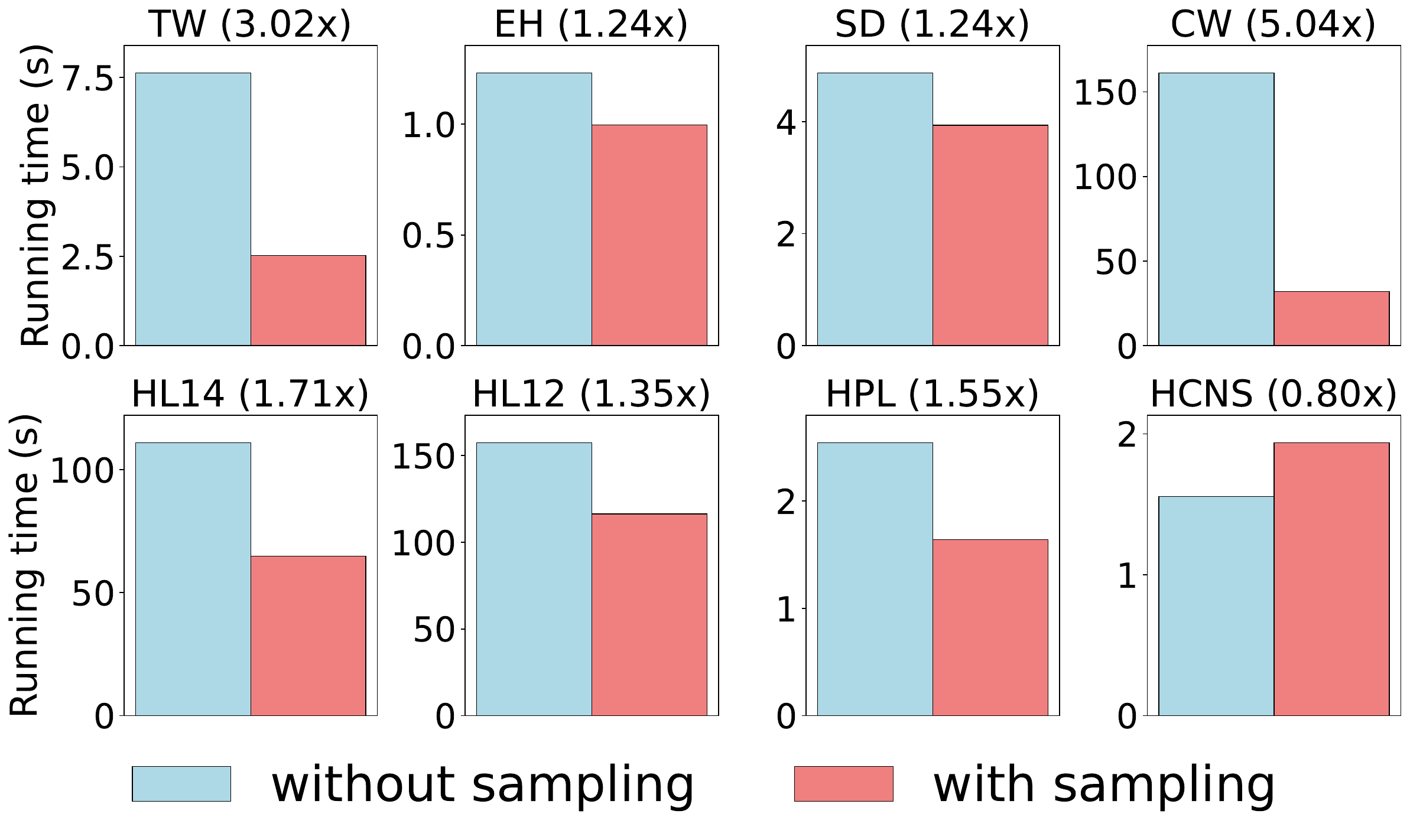}
  \caption{\textbf{Running time comparison w/ and w/o sampling.}
  The numbers at the title of each subfigure are the running time speedup of the sampling method over the non-sampling method.
  }\label{fig:exp-sampling}
\end{figure}

To make the effect of sampling clearer, here we show a separate figure to illustrate the improvement from sampling on the eight graphs that trigger sampling. 
We compare the running time of our algorithm with and without sampling. The results are shown in \cref{fig:exp-sampling}. 
The only graph whose performance is not improved by sampling is \HCNS{},
because sampling does not provide contention reduction for the graph,
while it adds overhead to the algorithm.
}

\section*{B. Maximum $k$-Core Subgraph}

As a fundamental component of many dense subgraph discovery algorithms, $k$-core decomposition has applications across a variety of domains. 
One frequently encountered task is identifying the maximum $k'$-core subgraph within a given graph, where all vertices have a degree of at least $k'$. Similar to the $k$-core decomposition process, this requires iterative peeling rounds to extract the $k'$-core subgraph. 
Our framework can be easily adapted to address this problem by modifying the peeling condition. Furthermore, local search techniques and sampling-based optimizations can be integrated to enhance performance across different graph types.

To the best of our knowledge, Galois~\cite{Nguyen2014} implemented the state-of-the-art algorithm for this subgraph finding problem.
We adapt our algorithm framework to the maximum $k-$core problem this problem and compared it with Galois.
In practice, subgraph finding is usually applied to dense social networks or web graphs,
which is critical for community detection and anomaly detection~\cite{zhang2017finding, kitsak2010identification,boguna2004models, giatsidis2011evaluating}.
We tested our implementation on two representative social networks, com-orkut~\cite{yang2015defining} and twitter~\cite{kwak2010twitter},
with $k$ value range from 16 to 2048.
Our algorithm outperforms Galois on both graphs, with a speedup of up to $3.5\times$.
The detailed results are shown in \cref{fig:subk}.
\revise{
\section*{C. Burdened Span Analysis}
As mentioned in the main paper, to better understand how VGC improves the burdened span and performance, we test the burdened span for our algorithm (with and without VGC), and compare it with \GBBS{}. 
In \cref{fig:app:burdened_span}, we present the speedup of our algorithm (with and without VGC) over \Julienne{} on burdened span. This is the same figure shown in our main paper. 
Here we further put a comparison on the running time of our algorithm (with and without VGC) over \Julienne{} in \cref{fig:app:burdened_span_speedup}. 
Comparing the trends, in most of the cases, the graphs that achieves the most significant speedup in burdened span also benefit the most from the running time. This indicates that a major performance gain of our algorithm over \Julienne{} is to reduce the burdened span using VGC.

\begin{figure}[t]
  \centering
  \includegraphics[width=\columnwidth]{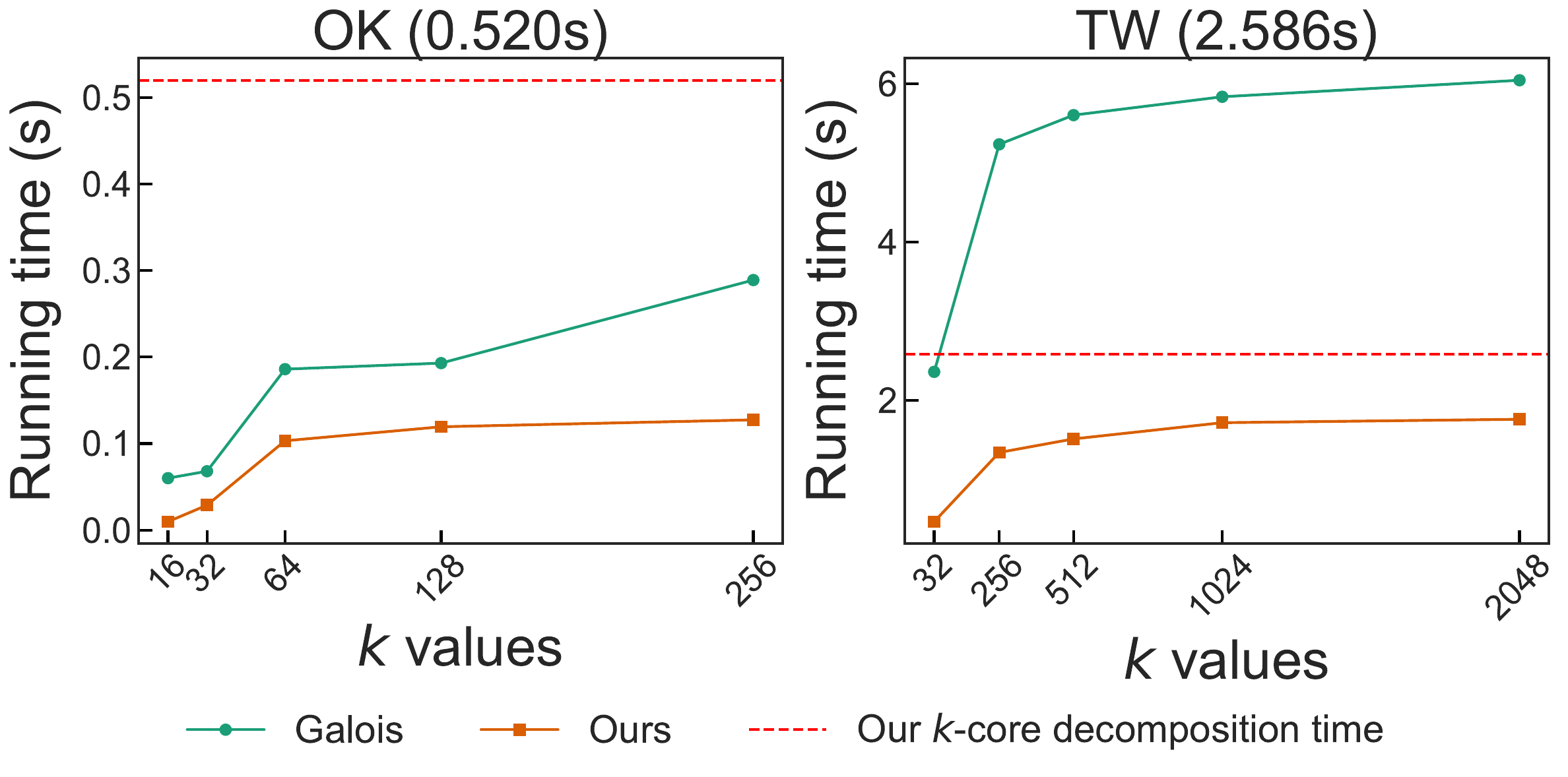}
  \caption{\textbf{Comparison of our adapted subgraph finding algorithm and Galois. 
  The $x$-axis is the $k$ value for subgraph finding.
  The running time at $y$-axis is the time in seconds.}
  The running time at the title is the $k$-core decomposition time of our algorithm in seconds.}\label{fig:subk}
\end{figure}

\begin{figure}[!t]
  \centering
  \includegraphics[width=\columnwidth]{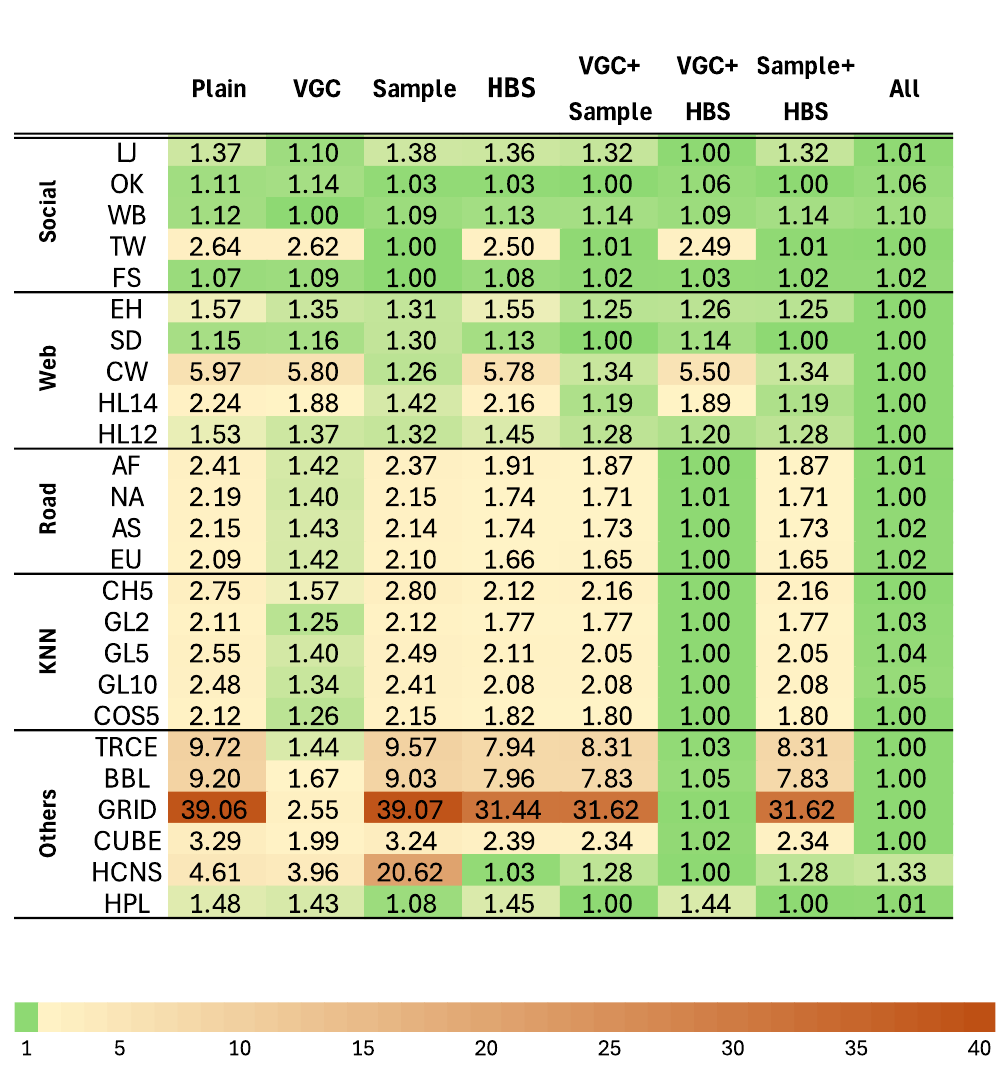}
  \caption{\label{fig:app:heatmap}\textbf{
    Heatmap of our 8 versions of algorithms (with and without VGC, sampling, and HBS). }
    The data is normalized to the minimum running time for each graph.
    The color gradient represents the percentile distribution of the running time data for the corresponding graph.
  }
\end{figure}

\section*{D. Evaluating All Combinations of the Three Techniques}

The results in \cref{table:combinations} shows the relative running time of eight combinations of the three proposed techniques (VGC, sampling, and HBS) on all graphs. The data is normalized to the minimum running time for each graph. We also give the raw data in \cref{table:combinations}. 
The heatmap in \cref{fig:app:heatmap} visualizes this data, 
with colors ranging from green to dark red representing relative running times on a 1 to 40 scale based on percentiles.
HBS, VGC, and sampling each contribute to performance improvements, though their impacts vary across graph types.
Among social graphs such as \TW and \CW, sampling plays a more critical role in improving the performance.
For sparse graphs such as \GL and \COS, VGC is the key technology to reduce the running time.
HBS provides consistent performance gains across all graphs and is particularly effective on sparse graphs. 
Interestingly, for each of graph, especially the adversarial cases, we observed that usually the good performance is provided by a specific combination of the techniques. For example, \CBC{} relies on both VGC and HBS to achieve a satisfactory performance; \SD{} requires both VGC and sampling; \HPL{} can achieve close-to-the-best performance just by sampling; and some graphs, such as \EH{}, cannot achieve the best performance without any of the three techniques. This indicates that the combinations of the three techniques in our solution is essential to guarantee the overall good performance and to handle various adversarial input instances. 

\clearpage

\begin{figure*}[b]
  \centering
    \includegraphics[width=\textwidth]{figures/charts/span_ratio.pdf}
  \caption{\textbf{Burdened span~\cite{he2010cilkview} speedup of our algorithm (with and without VGC) over \GBBS~\cite{dhulipala2017, gbbs2021} (green dotted line, always$=1$) on all graphs. Higher is better.}
  The bars are truncated at 18 for better visualization. The text on the bars are actual burdened span speedup.
  When no HBS is used, we uses 16 buckets as in the \Julienne{} implementation. 
  \label{fig:app:burdened_span}}
  \includegraphics[width=\textwidth]{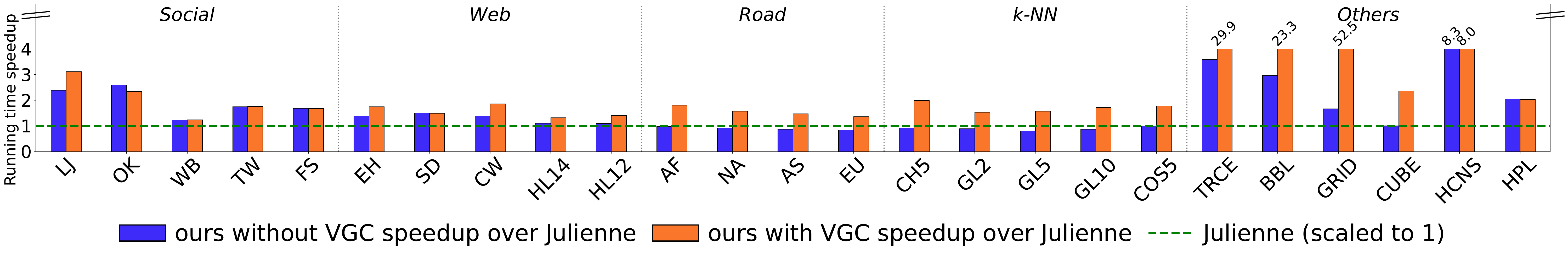}
  \caption{\textbf{Running time speedup of our algorithm (with and without VGC) over \GBBS~\cite{dhulipala2017, gbbs2021} (green dotted line, always$=1$) on all graphs. Higher is better.}
  The bars are truncated at 4 for better visualization. The text on the bars are actual running time speedup.
  When no HBS is used, we uses 16 buckets as in the \Julienne{} implementation. 
  \label{fig:app:burdened_span_speedup}}
\end{figure*} 

\begin{table*}[htbp] 
      \centering
      \small
      \vspace{-1em}

    \begin{tabular}{ccrrrr|rrrrrrrr|l}
          & \multicolumn{5}{c|}{\textit{\textbf{Graph Statistics}}} & \multicolumn{1}{c}{\multirow{2}[1]{*}{Plain}} & \multicolumn{1}{c}{\multirow{2}[1]{*}{VGC}} & \multicolumn{1}{c}{\multirow{2}[1]{*}{Sample}} & \multicolumn{1}{c}{\multirow{2}[1]{*}{HBS}} & \multicolumn{1}{c}{VGC+} & \multicolumn{1}{c}{VGC+} & \multicolumn{1}{c}{Sample+} & \multicolumn{1}{c|}{\multirow{2}[1]{*}{All}} & \multicolumn{1}{c}{\multirow{2}[1]{*}{\textit{\textbf{Notes}}}} \\
          & Name  & \multicolumn{1}{c}{$n$} & \multicolumn{1}{c}{$m$} & \multicolumn{1}{c}{$\maxcoreness$} & \multicolumn{1}{c|}{$\rho$} &       &       &       &       & \multicolumn{1}{c}{Sample} & \multicolumn{1}{c}{HBS} & \multicolumn{1}{c}{HBS} &       &  \\
\cmidrule{2-15}    \multicolumn{1}{c}{\multirow{5}[2]{*}{\begin{sideways}Social\end{sideways}}} & LJ    & 4.85M & 85.7M & 372   & 3,480 & .275  & .220  & .276  & .272  & .265  & \textbf{.200} & .265  & .203  & soc-LiveJournal1~\cite{backstrom2006group} \\
          & OK    & 3.07M & 234M  & 253   & 5,667 & .528  & .540  & .488  & .487  & .474  & .510  & .474  & .526  & com-orkut~\cite{yang2015defining} \\
          & WB    & 58.7M & 523M  & 193   & 2,910 & .934  & .831  & .902  & .937  & .946  & .913  & .946  & .935  & soc-sinaweibo~\cite{nr} \\
          & TW    & 41.7M & 2.41B & 2,488 & 14,964 & 7.15  & 7.09  & 2.71  & 6.77  & 2.74  & 6.73  & 2.74  & 2.72  & Twitter~\cite{kwak2010twitter} \\
          & FS    & 65.6M & 3.61B & 304   & 10,034 & 3.85  & 3.90  & 3.59  & 3.86  & 3.67  & 3.70  & 3.67  & 3.67  & Friendster~\cite{yang2015defining} \\
\cmidrule{2-15}    \multirow{5}[2]{*}{\begin{sideways}Web\end{sideways}} & EH    & 11.3M & 522M  & 9,877 & 7,393 & 1.25  & 1.07  & 1.04  & 1.23  & .996  & 1.00  & .996  & .795  & eu-host~\cite{webgraph} \\
          & SD    & 89.3M & 3.88B & 10,507 & 19,063 & 5.03  & 5.07  & 5.70  & 4.96  & 4.37  & 4.97  & 4.37  & 4.39  & sd-arc~\cite{webgraph} \\
          & CW    & 978M  & 74.7B & 4,244 & 106,819 & 171   & 166   & 36.1  & 165   & 38.3  & 157   & 38.3  & 28.6  & ClueWeb~\cite{webgraph} \\
          & HL14  & 1.72B & 124B  & 4,160 & 58,737 & 123   & 103   & 78.0  & 118   & 65.0  & 103   & 65.0  & 54.7  & Hyperlink14~\cite{webgraph} \\
          & HL12  & 3.56B & 226B  & 10,565 & 130,737 & 166   & 148   & 143   & 157   & 138   & 130   & 138   & 108.4 & Hyperlink12~\cite{webgraph} \\
\cmidrule{2-15}    \multirow{4}[2]{*}{\begin{sideways}Road\end{sideways}} & AF    & 33.5M & 88.9M & 3     & 189   & .372  & .219  & .366  & .294  & .288  & .154  & .288  & .155  & OSM Africa~\cite{roadgraph} \\
          & NA    & 87.0M & 220M  & 4     & 286   & .946  & .605  & .931  & .751  & .739  & .437  & .739  & .432  & OSM North America~\cite{roadgraph} \\
          & AS    & 95.7M & 244M  & 4     & 343   & 1.02  & .674  & 1.01  & .818  & .816  & .471  & .816  & .480  & OSM Asia~\cite{roadgraph} \\
          & EU    & 131M  & 333M  & 4     & 513   & 1.39  & .948  & 1.40  & 1.11  & 1.10  & .666  & 1.10  & .679  & OSM Europe~\cite{roadgraph} \\
\cmidrule{2-15}    \multirow{5}[2]{*}{\begin{sideways}\KNN\end{sideways}} & CH5   & 4.21M & 29.7M & 5     & 7     & .058  & .033  & .059  & .045  & .046  & .021  & .046  & .021  & Chem~\cite{fonollosa2015reservoir,wang2021geograph}, $k=5$ \\
          & GL2   & 24.9M & 65.3M & 2     & 12    & .223  & .133  & .224  & .187  & .187  & .106  & .187  & .109  & GeoLife~\cite{geolife,wang2021geograph}, $k=2$ \\
          & GL5   & 24.9M & 157M  & 5     & 42    & .306  & .168  & .299  & .253  & .246  & .120  & .246  & .125  & GeoLife~\cite{geolife,wang2021geograph}, $k=5$ \\
          & GL10  & 24.9M & 310M  & 10    & 16    & .380  & .206  & .370  & .320  & .319  & .154  & .319  & .162  & GeoLife~\cite{geolife,wang2021geograph}, $k=10$ \\
          & COS5  & 321M  & 1.96B & 2     & 23    & 4.33  & 2.58  & 4.38  & 3.71  & 3.68  & 2.04  & 3.68  & 2.04  & Cosmo50~\cite{cosmo50,wang2021geograph}, $k=5$ \\
\cmidrule{2-15}    \multirow{6}[2]{*}{\begin{sideways}Others\end{sideways}} & TRCE  & 16.0M & 48.0M & 2     & 1,839 & .638  & .095  & .628  & .521  & .545  & .067  & .545  & .066  & Huge traces~\cite{nr} \\
          & BBL   & 21.2M & 63.6M & 2     & 1,915 & .712  & .129  & .699  & .616  & .605  & .082  & .605  & .077  & Huge bubbles~\cite{nr} \\
          & GRID  & 100M  & 400M  & 2     & 50,499 & 11.0  & .718  & 11.0  & 8.86  & 8.91  & .284  & 8.91  & .282  & Huge grids \\
          & CUBE  & 1.00B & 6.0B  & 3     & 2,895 & 13.2  & 7.98  & 13.0  & 9.57  & 9.38  & 4.11  & 9.38  & 4.01  & Huge cubic \\
          & HCNS  & 0.1M  & 5.0B  & 50,000 & 50,000 & 6.96  & 5.98  & 31.1  & 1.56  & 1.94  & 1.51  & 1.94  & 2.01  & Huge Coreness \\
          & HPL   & 100M  & 1.20B & 3,980 & 6,297 & 2.58  & 2.50  & 1.89  & 2.52  & 1.75  & 2.52  & 1.75  & 1.77  & Huge scale-free \\
\cmidrule{2-15}    \end{tabular}%

  \caption{\textbf{Graph information and running time (in seconds) of the combinations of our three techniques (sample, VGC, and HBS).} 
  \textnormal{
    $n= |V|$, $m= |E|$.
    $\maxcoreness=$ maximum coreness value. 
    $\rho$ = the peeling complexity~\cite{dhulipala2017}, i.e., the number of subrounds using offline peeling. 
    }
      \label{table:combinations}
      }
    
    \end{table*}%
 
} 
}
\end{document}
\endinput